\documentclass[twocolumn,aps,prx,superscriptaddress,longbibliography,floatfix]{revtex4-2}

\usepackage{amsmath,amssymb,graphicx,booktabs,capt-of}
\usepackage{placeins}
\usepackage{float}
\usepackage[hidelinks]{hyperref}
\usepackage{xcolor}
\graphicspath{{figures/}}

\begin{document}

\title{Decoder Model Compatibility Provides Information beyond the Logical Gap under Drifting and Correlated Quantum Noise}

\author{Aaron C. Hoyt}
\affiliation{University of Washington, Seattle, Washington 98195, USA}
\affiliation{Pacific Northwest National Laboratory, Richland, Washington 99354, USA}

\author{Chunshu Wu}
\affiliation{Pacific Northwest National Laboratory, Richland, Washington 99354, USA}

\author{Shuwen Kan}
\affiliation{Pacific Northwest National Laboratory, Richland, Washington 99354, USA}
\affiliation{Department of Computer and Information Science, Fordham University, New York City, New York 10458, USA}

\author{Sean Garner}
\affiliation{University of Washington, Seattle, Washington 98195, USA}
\affiliation{Pacific Northwest National Laboratory, Richland, Washington 99354, USA}

\author{Avimita Chatterjee}
\affiliation{Lawrence Berkeley National Laboratory, Berkeley, California 94720, USA}

\author{Drew Rebar}
\affiliation{Pacific Northwest National Laboratory, Richland, Washington 99354, USA}

\author{Norman M. Tubman}
\affiliation{NASA Ames Research Center, Moffett Field, California 94035, USA}

\author{Katherine Klymko}
\affiliation{Lawrence Berkeley National Laboratory, Berkeley, California 94720, USA}

\author{Chenxu Liu}
\affiliation{Pacific Northwest National Laboratory, Richland, Washington 99354, USA}

\author{Ang Li}
\affiliation{University of Washington, Seattle, Washington 98195, USA}

\author{Samuel Stein}
\affiliation{Pacific Northwest National Laboratory, Richland, Washington 99354, USA}

\begin{abstract}
Quantum error-correcting decoders provide not only corrections but also confidence information about the logical-error risk of individual measurement records. This information can help characterize logical-error risk and guide postselection, where high-risk shots are rejected to improve fidelity. Its effectiveness, however, depends critically on the accuracy of the decoder's underlying noise model: when this model is mismatched to the physical device, confidence estimates such as the logical gap can become poorly calibrated. Such mismatch is difficult to avoid in real quantum hardware, where temporal drift, calibration uncertainty, and other nonstationary effects cause physical noise to deviate from a fixed decoder model. In this work, we show that this mismatch can itself provide useful information for postselection. We demonstrate on IBM quantum hardware that syndrome statistics drift across consecutive acquisition windows, causing detector histories to depart substantially from those predicted by the stationary detector error model (DEM) used for decoding. We observe a similar mismatch in independent Google surface-code memory experiments. We exploit this miscalibration by assigning additional rejection weight to detector histories that are unlikely under the assumed decoder model. For minimum-weight perfect matching (MWPM), the minimum weight $W$ provides a computationally inexpensive proxy for syndrome surprisal under the decoder model and therefore for identifying such unlikely detector histories. On held-out IBM repetition-code memories decoded with MWPM, our likelihood-aware postselection method reduces the retained logical-error probability (LEP) by $10.6\%$ at $d=7$ and $28.8\%$ at $d=9$, both at $15\%$ rejection. On openly available Google ten-round $d=5$ XZZX surface-code memory data, the method reduces the retained LEP by $75.4\%$ at $90\%$ rejection. Applied to Google's magic-state cultivation experiment with the tesseract decoder, it reduces tomography infidelity by $34.7\%$ with $30\%$ additional grafting-stage rejection, or increasing the accepted yield by $30.4\%$ at a target fidelity of $97.0\%$. Simulations under burst noise further show that the advantage persists as code distance and syndrome-extraction volume increase. These results show that decoder–hardware mismatch can be exploited as a signal of logical-error risk. Our method reuses quantities already produced during logical-gap decoding, requiring only a lightweight fitting step and no decoder modification or additional decoding passes, providing a practical route to improved reliability and yield under realistic nonstationary noise.

\end{abstract}

\maketitle

\section{Introduction}

Quantum error correction provides a route to fault-tolerant quantum computation by encoding logical information so that physical errors can be detected and corrected \cite{NielsenChuang2010}. Decoders use the resulting syndrome information to infer corrections, but they can also provide soft information about the reliability of those decisions, giving a record-dependent estimate of logical-error risk. This decoder confidence is useful whenever logical operations or prepared states can be accepted, rejected, or otherwise treated according to their inferred reliability. In particular, postselection can use decoder confidence to trade acceptance rate for logical fidelity, with applications ranging from logical-state preparation to retryable fault-tolerant primitives such as non-Clifford resource-state preparation and magic-state factories \cite{Bombin2024,Gidney2024,Rosenfeld2025}.

Decoder soft information provides a principled basis for making this tradeoff. Bombin et al.\ introduced the complementary logical gap as a confidence metric for fault-tolerant postselection and also proposed nested- and radial-gap rules that refine the discrete MWPM gap in their preparation geometry \cite{Bombin2024}. Those refinements address limitations of the approximate gap; here we focus on confidence miscalibration caused by mismatch between the detector error model and physical noise. Related exclusive decoders likewise reject difficult syndrome records \cite{Smith2024}. English et al.\ show that, when the decoder evaluates the true physical noise distribution, ranking records by their posterior logical confidence gives the optimal postselection rule at fixed acceptance \cite{English2025}. This result both motivates the logical gap as a confidence measure and exposes an important limitation: its optimality assumes that the physical noise distribution is accurately known by the decoder.

Several recent works have broadened how decoder confidence is constructed and used for postselection. One approach is to extract additional confidence by modifying or constraining the decoding problem. Xie et al.\ introduced argument reweighting, which performs additional decoding rounds under reweighted noise models to probe the robustness of the decoder's decision, while Wills et al.\ developed forced-gap postselection, in which the decoder is forced toward complementary logical outcomes and their likelihoods are compared \cite{Xie2026,Wills2026}. Other works have developed confidence metrics tailored to particular stages of fault-tolerant computation or to broader classes of codes. Staples et al.\ introduced the partial gap, which estimates the eventual logical gap of a resource state before it is consumed and enables postselection of intermediate fault-tolerant primitives \cite{Staples2026}. Lee et al.\ proposed cluster-based confidence metrics derived from cluster sizes and log-likelihood-ratio weights for general quantum LDPC codes, and benchmarked correction weight as a simpler postselection baseline \cite{Lee2026}. Dinc\u{a} et al.\ studied the swim distance, a computationally cheaper geometric confidence score derived from syndrome clusters, and compared it with the complementary gap \cite{Dinca2025}. Decoder confidence has also been explored experimentally and through learned models. Sales Rodriguez et al.\ used logical-gap-based postselection in an experimental neutral-atom magic-state-distillation protocol \cite{Rodriguez2025}, while Haug et al.\ applied a syndrome-only learned confidence score to those experimental data and showed that it can complement logical-gap filtering \cite{Haug2026}. Dentelski compared the learned confidence of a graph-neural-network decoder with the MWPM gap on identical simulated surface-code syndromes \cite{Dentelski2026}. These works develop alternative ways to construct or exploit decoder confidence. Here, we address a complementary question: how should a decoder's reported confidence be interpreted when the noise model used to compute it is mismatched to the physical device? For MWPM, we show that the existing minimum-cost output $W$ provides additional information about this model mismatch and can be incorporated using only a scalar coefficient selected on validation data.

The assumption that the decoder model matches the physical noise is central to the interpretation of its reported confidence. Experimental decoder models are necessarily approximate: they may be constructed from finite calibration data, inferred from measured syndrome statistics, or restricted to simplified model classes that permit efficient decoding \cite{Chen2022,Sivak2024,TakouGraphs2025}. Syndrome-based noise estimation and adaptive weighting offer one response \cite{Spitz2018,Wagner2021,BlumeKohout2025,Kobori2025,Bhardwaj2025,TakouCoherent2025}; here we study records evaluated by a frozen, potentially imperfect decoder. Moreover, the physical noise itself need not remain stationary. Superconducting qubits exhibit substantial temporal fluctuations in their energy-relaxation times, which have been associated with the spectral dynamics of microscopic two-level defects and with fluctuations in quasiparticle populations \cite{Carroll2022,Zhu2025}. Environmental disturbances can also produce correlated changes across an entire processor. Ionizing radiation can generate energetic phonons and nonequilibrium quasiparticles throughout a superconducting chip, producing simultaneous reductions in $T_1$ across many qubits \cite{Vepsalainen2020,Wilen2021,McEwen2022}. The thermal environment provides another source of variability: thermally excited quasiparticles contribute to qubit relaxation, while residual thermal photons in readout resonators can substantially increase dephasing \cite{Catelani2011,Wang2019}. Related common-mode effects arise in other hardware platforms. Trapped-ion processors can employ optical fields shared across many ions, with slow laser-frequency and intensity drifts contributing to gate errors \cite{Lu2019,ChenIon2023}; neutral-atom processors similarly use global Rydberg control and are sensitive to laser noise and intensity fluctuations \cite{Evered2023,Fromonteil2023}. A decoder calibrated at one time can therefore remain fixed while the physical distribution generating its syndrome records changes, potentially in a correlated manner across many qubits and operations.

Under such model mismatch, relative logical confidence and absolute model compatibility become distinct. The logical gap measures how strongly one logical sector is preferred over another, but not how well either sector explains the observed syndrome under the decoder model. Two records can therefore have the same reported logical gap even when one is typical under the assumed model and the other is from anomalous noise sources. Under an exact, correctly specified model this distinction carries no additional information about logical risk; under mismatch, the two records need not have the same physical probability of failure.

The distinction between relative logical confidence and decoder-model compatibility is particularly simple for MWPM. The two sector-constrained decoding costs determine both the complementary logical gap $\Delta$ and the smaller cost
\begin{equation}
    W=\min(C_0,C_1),
\end{equation}
which measures how costly even the decoder's preferred explanation is. In the ground-state approximation underlying MWPM, $W$ is a proxy for syndrome surprisal at fixed $\Delta$. Section~II places these quantities in a common likelihood framework and derives the response of the logical free-energy gap to model mismatch. The resulting expression shows that mismatch changes logical confidence only when it reweights the competing logical sectors differently, thereby identifying the condition under which absolute model compatibility can provide information beyond the reported gap.

This mechanism is isolated in Sec.~III using controlled models of drifting noise. Simulations of bursting noise show that a stationary detector error model can reproduce the time-averaged mechanism rates while failing to reproduce the joint syndrome distribution. An exactly enumerable surface-code circuit then makes it possible to track the physical logical confidence as the noise distribution departs from the frozen decoder model. Within the low-gap region, high-$W$ records acquire the largest increase in logical risk, directly linking model incompatibility to miscalibration of the reported gap. Circuit-level simulations at larger distance show that the resulting advantage of likelihood-aware postselection persists as the code distance and syndrome-extraction volume increase under burst noise.

The same structure appears in experimental data. Section~IV shows that time series IBM repetition code memory experiments exhibit persistent temporal variation in syndrome statistics and a substantially heavier high-$W$ tail than predicted by calibration-derived detector error models. In Sec.~V, incorporating $W$ into logical-gap postselection reduces the retained logical-error probability on held-out IBM data by $10.64\%$ $[8.35\%,16.18\%]$ at $d=7$ and $28.75\%$ $[26.84\%,30.20\%]$ at $d=9$ at $15\%$ rejection. The same construction reduces the retained logical-error probability by $75.4\%$ $[50.8\%,100\%]$ at $90\%$ rejection for ten-round $d=5$ Google XZZX surface-code memories. That section also extends the method to Google's tesseract-code magic-state cultivation experiment, where likelihood-aware postselection reduces tomography infidelity by $34.74\%$ $[27.21\%,42.39\%]$ at $30\%$ additional grafting-stage rejection, or increases accepted yield by $30.44\%$ $[11.73\%,42.04\%]$ at a target fidelity of $97\%$. Taken together, these results across distinct codes, decoder models, hardware platforms, and fault-tolerant tasks support a broader likelihood-aware approach in which absolute model compatibility supplements relative logical confidence when the assumed noise model is imperfect.

\section{Decoder confidence under model mismatch}

This section establishes the decoding formalism used throughout the paper. We first define the stabilizer, detector, and logical-sector structure of the memory experiments considered below. We then formulate maximum-likelihood decoding as a statistical-mechanical problem in which the relative likelihood of the two logical sectors determines the decoder confidence. This representation also makes the effect of model mismatch explicit: when the physical noise differs from the distribution assumed by the decoder, the reported logical gap need not equal the physical logical confidence. Finally, we specialize the construction to minimum-weight perfect matching (MWPM), where the logical gap and a second quantity measuring compatibility with the decoder model can both be obtained from the sector-constrained matching costs.

\subsection{Logical sectors and decoder confidence}

A stabilizer code is defined by a commuting subgroup $\mathcal S$ of the Pauli group, with code space given by the simultaneous $+1$ eigenspace
\begin{equation}
    \mathcal C=
    \left\{
    \lvert\psi\rangle :
    S\lvert\psi\rangle=\lvert\psi\rangle
    \;\; \forall\,S\in\mathcal S
    \right\}.
\end{equation}
Logical Pauli operators preserve $\mathcal C$ while acting nontrivially on the encoded information. Equivalently, they commute with every stabilizer but are not themselves stabilizers. For one encoded qubit, representatives $\overline X$ and $\overline Z$ may be chosen such that
\begin{equation}
    [\overline X,S]=[\overline Z,S]=0
    \quad \forall\,S\in\mathcal S,
    \qquad
    \{\overline X,\overline Z\}=0.
\end{equation}

For the distance-$d$ repetition code, the $Z$-basis memory has stabilizer generators and logical operators
\begin{equation}
    S_i=Z_iZ_{i+1},
    \qquad
    \overline Z=Z_1,
    \qquad
    \overline X=\prod_{i=1}^{d}X_i,
\end{equation}
with $i=1,\ldots,d-1$, while the $X$-basis memory follows by exchanging $X\leftrightarrow Z$. On a rotated surface-code patch, the bulk stabilizers are weight-four $X$- and $Z$-type plaquette operators, with reduced-weight checks at the boundaries. Logical operators may be chosen as strings connecting opposite boundaries,
\begin{equation}
    \overline Z=\prod_{q\in\gamma_Z}Z_q,
    \qquad
    \overline X=\prod_{q\in\gamma_X}X_q,
\end{equation}
where $\gamma_Z$ connects the two rough boundaries and $\gamma_X$ connects the two smooth boundaries. The two strings cross an odd number of times and therefore anticommute.

The syndrome in round $t$ is the vector $\mathbf s_t=(s_{1,t},\ldots,s_{n_s,t})$ of measured stabilizer outcomes, where $s_{a,t}\in\{0,1\}$ corresponds to stabilizer eigenvalue $(-1)^{s_{a,t}}$. A detector instead records an XOR of measurement outcomes whose parity is deterministic in the absence of faults. For a check measured in consecutive rounds,
\begin{equation}
    d_{a,t}=s_{a,t}\oplus s_{a,t-1}.
\end{equation}
Thus $\mathbf s_t$ specifies the stabilizer values at a particular time, whereas $d_{a,t}=1$ marks a change in one of those values and hence a detection event. Temporal-boundary detectors are defined analogously using the known initialization or the final data-qubit measurements. We denote the complete spacetime detector history by
\begin{equation}
    \mathbf d=(d_1,\ldots,d_m),
\end{equation}
and decode the full history jointly as a single decoding instance.

A physical fault history induces an effective Pauli error $\mathcal E$, while the decoder uses $\mathbf d$ to infer a recovery operator $\mathcal R(\mathbf d)$. Since the physical error and recovery are compatible with the same detector history, their product $\mathcal R(\mathbf d)\mathcal E$ has trivial syndrome and therefore acts within the code space. For one encoded qubit, this residual operation is equivalent modulo stabilizers to one of $\overline I$, $\overline X$, $\overline Z$, or $\overline Y$. A memory experiment prepares and measures a single logical observable $\overline L$, so these four classes reduce to two relevant logical sectors according to whether the residual operation preserves or flips $\overline L$:
\begin{equation}
    \lambda(\mathcal E,\mathbf d)=
    \begin{cases}
        0, & [\mathcal R(\mathbf d)\mathcal E,\overline L]=0,\\
        1, & \{\mathcal R(\mathbf d)\mathcal E,\overline L\}=0.
    \end{cases}
\end{equation}
For example, in a $\overline Z$ memory, the residual classes $\{\overline I,\overline Z\}$ preserve the measured logical value, whereas $\{\overline X,\overline Y\}$ flip it. Distinct physical fault histories can therefore generate the same detector history while belonging to different logical sectors.

The decoder maps $\mathbf d$ to a predicted sector $\widehat\lambda(\mathbf d)$. Let $\ell\in\{0,1\}$ denote whether the final logical measurement is preserved or flipped relative to the prepared logical value. Defining
\begin{equation}
    Y=\mathbf 1[\widehat\lambda(\mathbf d)\ne\ell],
\end{equation}
the logical-error probability is
\begin{equation}
    p_{\mathrm L}=\Pr(Y=1).
    \label{eq:logical-error-probability}
\end{equation}
Postselection retains only detector histories satisfying a confidence rule; we call the retained fraction the coverage and compare selectors at exactly matched coverage.

\subsection{Model mismatch and physical logical confidence}

For a fixed detector history $\mathbf d$, the recovery $R(\mathbf d)$ introduced above is fixed, while the effective Pauli error $E$ that generated the record remains unknown. In general, many errors are compatible with the same detector history. Each candidate error is assigned to a logical sector according to the action of the residual operator $R(\mathbf d)E$ on the measured logical observable $L$, as defined above. We denote the set of errors compatible with $\mathbf d$ and sector $\lambda$ by
\[
\Omega_\lambda(\mathbf d)
=
\left\{
E :
E\mapsto\mathbf d,\,
\lambda(E,\mathbf d)=\lambda
\right\}.
\]
Thus, the detector history and decoder recovery are held fixed, while $\Omega_\lambda(\mathbf d)$ contains the different unobserved physical errors that could have generated the record and leave the same logical action after recovery.

For the single-logical-qubit memory experiments considered here, errors within a fixed sector can differ by stabilizers and by the measured logical operator $L$. To see this, choose a reference error $E_\lambda(\mathbf d)\in\Omega_\lambda(\mathbf d)$. Any other error in the same sector can be written, up to an irrelevant Pauli phase, as
\[
E(\mathbf n;\mathbf d,\lambda)
=
E_\lambda(\mathbf d)
L^{n_0}
\prod_{a=1}^{r}S_a^{n_a},
\]
where $S_1,\ldots,S_r$ are stabilizer generators and $n_a\in\{0,1\}$. Multiplication by a stabilizer changes the physical error representative without changing its detector history or residual logical action. Multiplication by the measured logical operator $L$ also leaves the detector history unchanged and does not change the sector, because it preserves the measured logical observable. For example, in a $Z$ memory the residual classes $\{I,\overline Z\}$ belong to the preserving sector, while $\{\overline X,\overline Y\}$ belong to the flipping sector.

For notational convenience, define $G_0=L$ and $G_a=S_a$ for $a=1,\ldots,r$, so that the compatible errors within a sector are generated by the binary variables $\mathbf n=(n_0,\ldots,n_r)$. We associate each binary variable with an auxiliary Ising spin
\[
\sigma_a=(-1)^{n_a}\in\{-1,+1\},
\]
and write $\boldsymbol{\sigma}=(\sigma_0,\ldots,\sigma_r)$. A spin configuration $\boldsymbol{\sigma}$ therefore labels one effective Pauli error compatible with the fixed detector history and logical sector. These spins are auxiliary variables used to enumerate the possible error configurations; they are neither physical qubit spins nor measured detector bits.

The likelihood assigned by the decoder to these configurations can be represented by an effective Ising Hamiltonian
\begin{equation}
    \begin{aligned}
        H_{0,\lambda}(\boldsymbol\sigma;\mathbf d)
        &=
        c_\lambda
        -\sum_a h_a\sigma_a
        -\sum_{a<b}J_{ab}\sigma_a\sigma_b
        \\
        &\quad
        -\sum_{a<b<c}K_{abc}\sigma_a\sigma_b\sigma_c
        -\cdots .
    \end{aligned}
    \label{eq:ising-decoder-main}
\end{equation}
The coefficients are determined by the noise model assumed by the decoder, with their dependence on the fixed detector history and logical sector left implicit. The one-body, pairwise, and higher-order terms describe how the decoder-assigned likelihood varies across the physical error configurations compatible with $\mathbf d$ and $\lambda$.

Let $q_0(E,\mathbf d)$ denote the joint likelihood assigned to an error configuration and detector history by the decoder's reference noise model. The likelihood of logical sector $\lambda$ is obtained by summing over all errors compatible with the observed detector history and that sector,
\begin{equation}
    Z_\lambda(\mathbf d)
    =
    \sum_{E\in\Omega_\lambda(\mathbf d)}
    q_0(E,\mathbf d)
    =
    \sum_{\boldsymbol\sigma}
    e^{-H_{0,\lambda}(\boldsymbol\sigma;\mathbf d)}.
    \label{eq:sector-partition-main}
\end{equation}
The sum therefore does not vary the observed detector history: it marginalizes over the different unobserved physical errors that could have produced that same record within sector $\lambda$.

The corresponding sector free energy is
\begin{equation}
    F_\lambda(\mathbf d)
    =
    -\log Z_\lambda(\mathbf d).
\end{equation}
The posterior probability assigned by the decoder to logical sector $\lambda$ is therefore
\begin{equation}
    q(\lambda\mid\mathbf d)
    =
    \frac{Z_\lambda(\mathbf d)}
    {Z_0(\mathbf d)+Z_1(\mathbf d)}.
\end{equation}
Maximum-likelihood decoding selects the sector with the lower free energy.

The signed exact logical free-energy gap is
\begin{equation}
    g_0(\mathbf d)
    =
    F_1(\mathbf d)-F_0(\mathbf d)
    =
    \log\frac{Z_0(\mathbf d)}{Z_1(\mathbf d)}.
\end{equation}
For the two logical sectors relevant to a memory experiment, the conditional logical-error probability under the decoder model is
\begin{equation}
    \Pr(\mathrm{fail}\mid\mathbf d)
    =
    \frac{1}{1+e^{|g_0(\mathbf d)|}}.
    \label{eq:exact-gap-confidence}
\end{equation}
Thus, when the decoder model matches the physical noise and the sector partition functions are evaluated exactly, $|g_0|$ completely determines the conditional logical risk.

The total probability assigned by the decoder to the observed detector history is proportional to the combined weight of the two sectors,
\begin{equation}
    q(\mathbf d)
    \propto
    Z_0(\mathbf d)+Z_1(\mathbf d),
\end{equation}
so its surprisal is
\begin{equation}
    A_0(\mathbf d)
    =
    -\log q(\mathbf d).
\end{equation}
The logical gap and the surprisal therefore quantify different properties of the decoder model: $|g_0|$ measures the relative preference between the two possible logical outcomes, whereas $A_0$ measures how well the observed detector history itself is supported by the assumed noise model.

This distinction becomes important when the physical noise differs from the model used by the decoder. We represent such a mismatch by perturbing the effective Hamiltonian,
\begin{equation}
    H_\epsilon
    =
    H_0+\epsilon V,
    \label{eq:mismatched-hamiltonian}
\end{equation}
where $\epsilon=0$ corresponds to the decoder model and $V$ describes the change in the physical error distribution. For a fixed detector history and logical sector, $V$ may alter the existing fields and couplings or introduce interactions absent from the reference Hamiltonian. The corresponding sector partition functions can be written
\begin{equation}
    Z_{\lambda,\epsilon}(\mathbf d)
    =
    Z_\lambda(\mathbf d)
    \left\langle
    e^{-\epsilon V}
    \right\rangle_{\lambda,\mathbf d},
\end{equation}
where $\langle\cdot\rangle_{\lambda,\mathbf d}$ denotes an expectation under the reference decoder model conditioned on $\mathbf d$ and sector $\lambda$. The physical logical gap therefore satisfies
\begin{equation}
    g_\epsilon(\mathbf d)-g_0(\mathbf d)
    =
    \log
    \left\langle e^{-\epsilon V}\right\rangle_{0,\mathbf d}
    -
    \log
    \left\langle e^{-\epsilon V}\right\rangle_{1,\mathbf d}.
    \label{eq:exact-gap-response}
\end{equation}

Model mismatch therefore changes the logical confidence only when it reweights the two sectors differently. A perturbation that affects both sectors equally may make a detector history much less likely without changing its logical odds. Absolute model compatibility can consequently identify records for which the frozen decoder model is poorly supported, but reduced compatibility alone does not imply increased logical risk; the mismatch must also distinguish between the competing logical sectors.

The complete cumulant expansion and its application to the oriented confidence used in the exact circuit experiment are given in Appendix~\ref{app:statistical-mechanical-decoding}.

\subsection{MWPM gap and model compatibility}

The preceding construction is exact but requires a sum over all error configurations compatible with each logical sector. MWPM replaces each sector partition function by its minimum-cost configuration, providing a ground-state approximation to the same statistical-mechanical problem. For an independent graphlike DEM, fault mechanism $e$ occurs with probability $p_e$ and is assigned matching weight
\begin{equation}
    w_e
    =
    \log\frac{1-p_e}{p_e}.
\end{equation}
A fault configuration $x=(x_1,\ldots,x_{N_e})$, with $x_e\in\{0,1\}$ indicating whether mechanism $e$ occurs, has additive cost
\begin{equation}
    C(x)
    =
    \sum_e w_e x_e.
\end{equation}

For a fixed detector history $\mathbf d$, let
\begin{equation}
    C_\lambda(\mathbf d)
    =
    \min_{x\rightarrow(\mathbf d,\lambda)}
    C(x)
\end{equation}
denote the minimum matching cost constrained to logical sector $\lambda$. In practice, these are the costs of the lowest-weight explanations of the same detector history in the two competing logical sectors. Their difference defines the complementary MWPM gap
\begin{equation}
    \Delta(\mathbf d)
    =
    \left|
    C_1(\mathbf d)-C_0(\mathbf d)
    \right|,
\end{equation}
while their minimum defines
\begin{equation}
    W(\mathbf d)
    =
    \min\!\left[
    C_0(\mathbf d),
    C_1(\mathbf d)
    \right].
    \label{eq:gap-weight-definitions}
\end{equation}
The two quantities have different interpretations. $\Delta$ measures how much more costly the competing logical explanation is than the decoder's preferred explanation and is therefore the MWPM analogue of relative logical confidence. By contrast, $W$ measures how costly even the preferred explanation is under the assumed detector error model and therefore probes absolute model compatibility.

The connection to the exact likelihood formalism follows by retaining only the minimum-energy configuration in each logical sector,
\begin{equation}
    Z_\lambda(\mathbf d)
    \simeq
    e^{-C_\lambda(\mathbf d)},
\end{equation}
up to terms independent of $\lambda$ and $\mathbf d$. Consequently,
\begin{equation}
    |g_0(\mathbf d)|
    \simeq
    \Delta(\mathbf d),
\end{equation}
so the complementary MWPM gap is the ground-state approximation to the exact logical free-energy gap.

The same approximation gives the detector-history surprisal
\begin{equation}
    -\log q(\mathbf d)
    \simeq
    W(\mathbf d)
    -
    \log\!\left[
    1+e^{-\Delta(\mathbf d)}
    \right]
    +
    \mathrm{const}.
    \label{eq:mwpm-surprisal-main}
\end{equation}
At fixed $\Delta$, the approximate surprisal therefore increases monotonically with $W$. The minimum sector cost $W$ is thus a ground-state proxy for how poorly the detector history is supported by the decoder model, while $\Delta$ quantifies the decoder's relative preference between the two logical sectors.

This provides the practical construction used throughout the remainder of the paper. A standard MWPM decoder already identifies the preferred minimum-cost correction; by additionally evaluating the minimum-cost correction in the complementary logical sector, one obtains both $C_0$ and $C_1$ and hence the pair $(\Delta,W)$. Under a correctly specified model, $\Delta$ contains the relevant logical-confidence information. Under model mismatch, however, the physical logical gap can depart from the gap reported by the frozen decoder. We therefore use $W$ alongside $\Delta$ to distinguish detector histories with similar reported logical ambiguity but different compatibility with the assumed noise model. The controlled models and experimental data below test when this additional coordinate improves postselection.

\section{Controlled model mismatch}

We now turn the two decoder coordinates into a postselection rule and test the mechanism that can make likelihood informative beyond the logical gap. The controlled studies separate three statements: the physical distribution can depart from the decoder model; that mismatch can change the true logical confidence; and the MWPM surprisal proxy can identify the records on which the change is dangerous.

\subsection{Likelihood-aware postselection}

For every simulation and hardware experiment, we first map each decoder coordinate to its cumulative probability under a reference distribution and rank records with
\begin{equation}
  S_\alpha(s)
  =1-\widehat F_\Delta[\Delta(s)]
  +\alpha\widehat F_W[W(s)].
  \label{eq:cdf-score}
\end{equation}
Here $\widehat F_\Delta$ and $\widehat F_W$ are mid-CDFs estimated from the reference model or from training data, as specified for each experiment in Appendix~\ref{app:common-coverage}. Explicitly,
\begin{equation}
  \widehat F_X(x)
  =
  \frac{1}{N}
  \sum_{i=1}^{N}
  \left[
    \mathbf{1}(X_i<x)
    +\frac{1}{2}\mathbf{1}(X_i=x)
  \right],
  \label{eq:mid-cdf}
\end{equation}
where the probabilities are empirical frequencies in the reference sample. This normalization puts the two coordinates on the same percentile scale across circuits. The first term ranks small gaps as dangerous, while the second raises the rejection priority of records in the high-$W$ tail; larger $S_\alpha$ is rejected first, and gap-only postselection is recovered at $\alpha=0$. The coefficient $\alpha$ may depend on the desired coverage and is selected using development and validation data only, except where a fixed value is stated explicitly. The coefficient, CDFs, and threshold are then frozen before held-out evaluation. Gap-only and gap+$W$ selectors are compared at exactly matched retained-shot counts, so an apparent improvement cannot arise from rejecting more data. When $\alpha$ is selected separately at each coverage, the plotted curve comprises separately validated operating points and need not correspond to nested accepted sets from one fixed ranking. Data split definitions are given in Appendix~\ref{app:common-coverage}.

\subsection{Breakdown of a stationary detector model}

A conventional detector error model assumes that a fixed set of error mechanisms occurs independently with stationary probabilities. We first demonstrate a limitation of this description: time-varying physical noise can generate detector statistics that cannot be represented by any single stationary independent-event DEM. We study this effect using Stim's rotated surface-code memory circuit~\cite{Gidney2021Stim} with distance $d=5$ and $r=5$ syndrome-extraction rounds. The circuit uses a homogeneous circuit-level Pauli noise model in which a common probability $p$ sets single- and two-qubit depolarizing noise after Clifford gates, depolarizing noise on data qubits before each syndrome-extraction round, and basis-appropriate flips after reset and before measurement.

The stationary reference circuit has noise strength $p_0=0.005$. To introduce controlled shot-to-shot variation, each shot is assigned a latent state $Z\in\{\mathrm{quiet},\mathrm{high}\}$, with a fraction $\rho=0.1$ of shots drawn from the high-noise state. We set
\begin{equation}
  p^{\mathrm{high}}=\xi p_0,
  \qquad
  p^{\mathrm{quiet}}
  =\frac{1-\rho\xi}{1-\rho}p_0,
  \label{eq:drifting-circuit-mixture}
\end{equation}
so that
\begin{equation}
  (1-\rho)p^{\mathrm{quiet}}
  +\rho p^{\mathrm{high}}
  =p_0.
  \label{eq:drifting-circuit-mean}
\end{equation}

Increasing the burst multiplier $\xi$ therefore strengthens the temporal variation while leaving the average error probability at every noisy circuit location unchanged. Conditioned on $Z$, each shot is generated by an ordinary stationary Stim circuit, but all noisy locations within that shot share the same value of $p$. After the latent state is ignored, this common fluctuation induces correlations between fault events and can alter the joint detector distribution even though the average circuit-level noise strength remains fixed.

\begin{figure*}[t]
  \centering
  \includegraphics[width=0.92\textwidth,trim=4pt 4pt 4pt 0pt,clip]{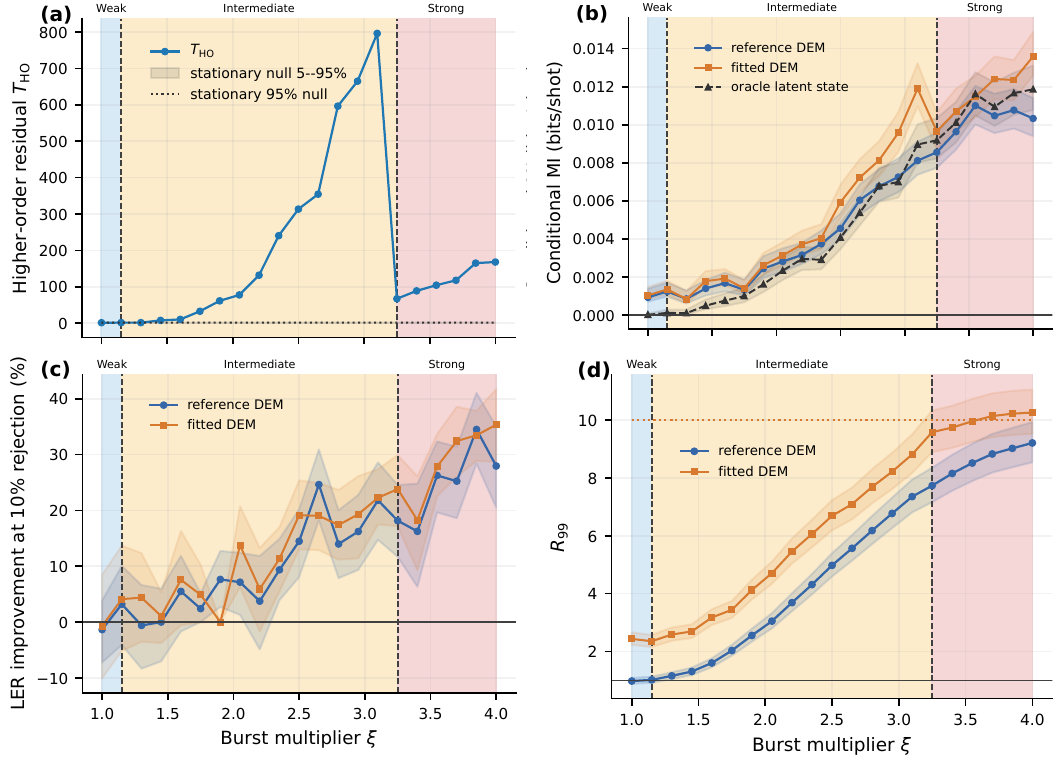}
  \caption{Controlled shot-to-shot variation in the $d=5$, $r=5$ circuit-level model. (a) Higher-order residual $T_{\mathrm{HO}}$ and its stationary null. (b) Conditional mutual information $I(Y;W\mid\Delta)$ for the reference and fitted decoders, together with the oracle information $I(Y;Z\mid\Delta)$ carried by the latent noise state. (c) Relative LEP improvement of gap+$W$ over gap-only postselection at matched coverage. (d) High-$W$ tail ratio relative to stationary samples from the corresponding decoder model. Vertical lines mark the detector-only boundaries $\xi_1$ and $\xi_2$; panel (a) shows the stationary-null 5th--95th percentile range and 95th-percentile threshold, while pointwise 95\% bands in panels (b), (c), and (d) use test-shot bootstrap, paired test-shot bootstrap, and Wilson ratio intervals, respectively.}
  \label{fig:controlled-drift}
\end{figure*}

We compare two stationary graphlike DEMs throughout the sweep. The \emph{reference DEM} is constructed once from the stationary $p=p_0$ circuit and held fixed, whereas the \emph{fitted DEM} is inferred separately at each $\xi$ from the detector statistics of the quiet/high-noise mixture. Let $d_i\in\{0,1\}$ indicate whether detector $i$ fires and let $\langle\cdot\rangle$ denote an average over fitting shots. Under the independent graphlike assumption, the bulk-edge probabilities are inferred from the measured one- and two-detector firing frequencies as~\cite{Bhardwaj2025,TakouGraphs2025}
\begin{equation}
  \widehat p_{ij}
  =\frac{1}{2}
  -\sqrt{
  \frac{1}{4}
  -
  \frac{
  \langle d_i d_j\rangle-\langle d_i\rangle\langle d_j\rangle
  }{
  1-2(\langle d_i\rangle+\langle d_j\rangle)+4\langle d_i d_j\rangle
  }},
  \label{eq:main-bulk-edge-inversion}
\end{equation}
with boundary-edge probabilities
\begin{equation}
  \widehat p_{ii}
  =
  \frac{1}{2}
  +
  \frac{\langle d_i\rangle-1/2}
  {\prod_{j\ne i}(1-2\widehat p_{ij})}.
  \label{eq:main-boundary-edge-inversion}
\end{equation}

When these equations return admissible mechanism probabilities, they directly define the fitted stationary DEM. Here, admissible means ordinary probabilities $0\leq\widehat p_e\leq1$ that also satisfy the graphlike DEM parameterization used here, $0\leq\widehat p_e<1/2$. The first value of $\xi$ for which this unconstrained inversion produces at least one $\widehat p_e$ outside that range defines $\xi_2$; beyond that point, the solution is discarded and a fit constrained to $0\leq\widehat p_e<1/2$ is used for the remaining diagnostics.

A successful low-order fit does not imply that the full detector distribution is described correctly. We therefore test the fitted DEM on connected three- and four-detector parity moments that were not used in the fit. For a detector subset $A$, the measured parity moment is
\begin{equation}
  m_A
  =
  \left\langle
  \prod_{i\in A}(1-2d_i)
  \right\rangle ,
  \label{eq:main-higher-order-moment}
\end{equation}
while the fitted independent DEM predicts
\begin{equation}
  m_A^{\mathrm{fit}}
  =
  \prod_e
  (1-2\widehat p_e)^{|A\cap E_e|\bmod 2},
  \label{eq:main-higher-order-prediction}
\end{equation}
where $E_e$ is the detector support of error mechanism $e$. The overlap $A\cap E_e$ contains the detectors shared by the tested subset and mechanism $e$, so $|A\cap E_e|$ is the number of shared detectors. Taking this number modulo two gives $0$ for an even overlap and $1$ for an odd overlap. Consequently, a mechanism contributes a factor only when it overlaps $A$ on an odd number of detectors. The product form follows from the independent-mechanism construction of a graphlike DEM; a derivation is given in Appendix~\ref{app:controlled-drift-methods}. The higher-order (HO) discrepancy between the measured and fitted moments is summarized by
\begin{equation}
  T_{\mathrm{HO}}
  =
  \frac{1}{|\mathcal H|}
  \sum_{A\in\mathcal H}
  \left(
  \frac{m_A-m_A^{\mathrm{fit}}}{\sigma_A}
  \right)^2,
  \label{eq:main-higher-order-statistic}
\end{equation}
  where $\mathcal H$ contains the tested connected three- and four-detector subsets and $\sigma_A$ is the corresponding stationary-null uncertainty.

To calibrate this statistic, we generate stationary records from the fitted DEM and repeat the complete fit-and-test procedure. We define $\xi_1$ as the first member of the first pair of consecutive $\xi$ values for which the observed $T_{\mathrm{HO}}$ exceeds the 95th percentile of this stationary null. The two detector-only boundaries $\xi_1$ and $\xi_2$ therefore mark, respectively, the first statistically resolved higher-order mismatch and the later point at which even the low-order inversion ceases to admit a stationary DEM with all mechanism probabilities in $0\leq\widehat p_e<1/2$.

The three regimes in Fig.~\ref{fig:controlled-drift}(a) make the breakdown explicit. For $\xi<\xi_1$, the temporal variation remains statistically compatible with a stationary independent DEM. In the intermediate region, $\xi_1\leq\xi<\xi_2$, the one- and two-detector statistics still yield a fitted DEM with every mechanism probability in $0\leq\widehat p_e<1/2$, yet the held-out higher-order moments reveal that this model does not reproduce the full detector distribution. Thus a stationary DEM can be fit successfully even when it is not a good model of the underlying process. Once $\xi\geq\xi_2$, at least one inferred mechanism probability lies outside $0\leq\widehat p_e<1/2$.

This mismatch also appears directly in the decoder likelihoods. As shown in Fig.~\ref{fig:controlled-drift}(d), the high-$W$ tail grows relative to stationary samples from the corresponding decoder model, indicating an increasing population of detector records that are unusually costly under the assumed stationary DEM. Refitting the DEM absorbs part of this change but does not remove it, consistent with the higher-order correlations that remain outside the stationary independent model class.

The excess cost is not merely an anomaly in the syndrome distribution. Fig.~\ref{fig:controlled-drift}(b) shows that $I(Y;W\mid\Delta)$ becomes nonzero as the mismatch grows, so $W$ carries information about the logical outcome $Y$ that is not already contained in the reported gap $\Delta$. The oracle quantity $I(Y;Z\mid\Delta)$ gives the corresponding information carried by the true quiet/high state and shows that the hidden temporal variation itself contains logical information beyond the gap.

The additional information has a direct operational consequence in Fig.~\ref{fig:controlled-drift}(c), where incorporating $W$ into the postselection score reduces the retained logical-error probability at fixed coverage relative to gap-only selection. Taken together, the controlled experiment shows that temporal mismatch can make a stationary DEM fail to capture the full detector distribution, broaden the high-$W$ likelihood tail, and leave logical-risk information in $W$ that is not contained in the reported gap.

Having established the limitations of the conventional stationary DEM and the emergence of additional information beyond its reported gap, we next examine the mechanism exactly. The following subsection tracks the physical logical confidence as the noise distribution departs from the frozen decoder model and shows how the decoder's logical gap becomes miscalibrated.

\subsection{Model mismatch and logical-risk miscalibration}

The previous subsection showed that shot-to-shot noise variation can produce detector statistics that are not captured by a stationary independent DEM. We now ask how this mismatch changes the logical confidence associated with a frozen reference model and why the absolute MWPM cost $W$ can become informative beyond the reported gap. To isolate the effect of model mismatch from the ground-state approximation made by MWPM, we first compute the logical confidence exactly from the sector probabilities of a small, enumerable circuit, and then compare the resulting risk response with both the exact and MWPM gaps. We use the same Stim circuit-level noise model at distance $d=5$ with $r=1$ syndrome-extraction round.

The frozen reference model is constructed from a quiet circuit with $p_0=0.005$. Let $q_\lambda(d)$ denote the joint probability of detector record $d$ and logical sector $\lambda\in\{0,1\}$ under this reference circuit, and let $q_\lambda^{\mathrm{hot}}(d)$ denote the corresponding probability under a hot circuit with $p_{\mathrm{hot}}=0.02$. Because the circuit is exactly enumerable, these sector probabilities can be evaluated without the MWPM ground-state approximation. We introduce the hot-shot fraction $\theta$ directly through
\begin{equation}
\begin{aligned}
  p_{\theta,\lambda}(d)
  &=
  (1-\theta)q_\lambda(d)
  +\theta q_\lambda^{\mathrm{hot}}(d),\\
  &\qquad 0\leq\theta\leq1.
\end{aligned}
  \label{eq:affine-mixture}
\end{equation}
Thus $\theta=0$ is the stationary distribution assumed by the frozen reference model, while increasing $\theta$ replaces an increasing fraction of quiet shots by hot shots. The interpolation acts on complete shot distributions, so every noisy circuit location within a sampled shot is generated from the same quiet or hot circuit model.

Let $\widehat\lambda(d)$ denote the logical sector selected by the frozen MWPM decoder. We orient the exact physical log-odds toward this fixed decision,
\begin{equation}
  h_\theta(d)
  =
  \log
  \frac{p_{\theta,\widehat\lambda(d)}(d)}
       {p_{\theta,1-\widehat\lambda(d)}(d)}.
  \label{eq:oriented-confidence}
\end{equation}
Importantly, $h_\theta(d)$ is not the MWPM gap: its magnitude is obtained from the exactly enumerated sector probabilities. At $\theta=0$, $|h_0(d)|=|\log[q_0(d)/q_1(d)]|$ is the exact logical gap of the quiet reference model, whereas for $\theta>0$, $|h_\theta(d)|$ is the exact logical gap of the physical quiet--hot mixture. The MWPM decoder enters here only through the orientation $\widehat\lambda(d)$ used to ask how reliable its fixed decision remains as the physical distribution moves away from the reference model.

The first-order response of this exact confidence at the reference model is
\begin{equation}
\begin{aligned}
  a(d)
  &\equiv
  \left.
  \frac{\partial h_\theta(d)}{\partial\theta}
  \right|_{\theta=0}\\
  &=
  \frac{q_{\widehat\lambda(d)}^{\mathrm{hot}}(d)}
       {q_{\widehat\lambda(d)}(d)}
  -
  \frac{q_{1-\widehat\lambda(d)}^{\mathrm{hot}}(d)}
       {q_{1-\widehat\lambda(d)}(d)}.
\end{aligned}
  \label{eq:exact-mixture-gap-response}
\end{equation}

Eq.~\eqref{eq:exact-mixture-gap-response} shows that model mismatch changes the logical confidence only when the two logical sectors respond differently. A detector record can become much more or less probable as hot shots are introduced without changing its logical odds if both sectors are reweighted by the same relative amount. This is the finite-distribution form of the sector-differential free-energy response derived in Appendix~\ref{app:statistical-mechanical-decoding}.

The conditional probability that the frozen MWPM decision is wrong under the physical mixture is
\begin{equation}
  r_\theta(d)
  =
  \frac{1}{1+e^{h_\theta(d)}},
\end{equation}
and its first-order response is
\begin{equation}
  \dot r(d)
  \equiv
  \left.
  \frac{\partial r_\theta(d)}{\partial\theta}
  \right|_{\theta=0}
  =
  -r_0(d)[1-r_0(d)]a(d).
  \label{eq:exact-mixture-risk-response}
\end{equation}
Positive $\dot r(d)$ therefore identifies detector records for which introducing hot shots increases the physical probability that the frozen decoder decision is incorrect.

\begin{figure}[t!]
  \centering
  \includegraphics[width=\columnwidth]{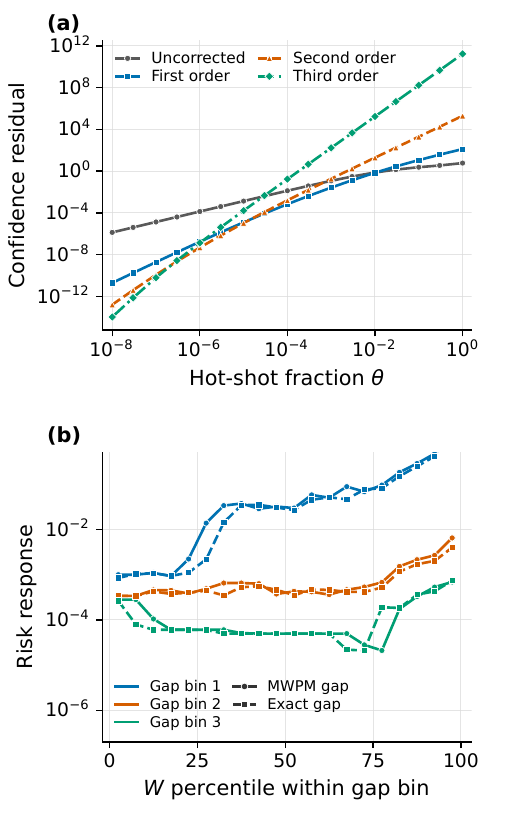}
  \caption{Exact $d=5$, $r=1$ response to increasing hot-shot fraction $\theta$. (a) Confidence residual between the exact physical log-odds $h_\theta$ of the quiet--hot mixture and approximations obtained by expanding about the exact quiet-model confidence $h_0$. The uncorrected curve therefore measures the miscalibration of the exact frozen-model gap itself as the physical distribution departs from the reference model; the first-, second-, and third-order curves successively incorporate the response to this mismatch. Residuals are reference-model-probability-weighted RMS values. (b) The mean risk response $\left.\partial r_\theta(d)/\partial\theta\right|_{\theta=0}$. Records are grouped into 20 within-gap $W$ percentile bins for each of the three lowest gap bins. Solid circles use the MWPM gap to define the gap bins, while dashed squares use the exact reference gap $|\log[q_0(d)/q_1(d)]|$. Both use the same MWPM minimum cost $W$ and exact physical risk response. Positive risk response means that increasing the hot-shot fraction makes the frozen decoder decision less reliable. The curves are computed by exact enumeration, so no sampling intervals are shown.}
  \label{fig:gap-correction-residuals}
\end{figure}

Fig.~\ref{fig:gap-correction-residuals}(a) directly shows the miscalibration produced by model mismatch even when the logical gap of the reference model is evaluated exactly. At $\theta=0$, the quiet-model sector probabilities give the exact logical confidence for the physical distribution. As $\theta$ increases, however, the true confidence becomes $h_\theta(d)$, while the frozen model continues to report $h_0(d)$. The uncorrected residual therefore measures the discrepancy between the exact logical gap inferred from the frozen quiet model and the exact logical gap of the actual quiet--hot mixture. Including the first-order response from Eq.~\eqref{eq:exact-mixture-gap-response} removes the leading $O(\theta)$ error and leaves an $O(\theta^2)$ residual, while successive orders further improve the local approximation. The expansion is local, so detector records with large relative changes between the quiet and hot distributions can limit the accuracy of a finite-order truncation at larger $\theta$.

\begin{figure*}[t!]
  \centering
  \includegraphics[width=\textwidth,trim=0 15pt 0 0,clip]{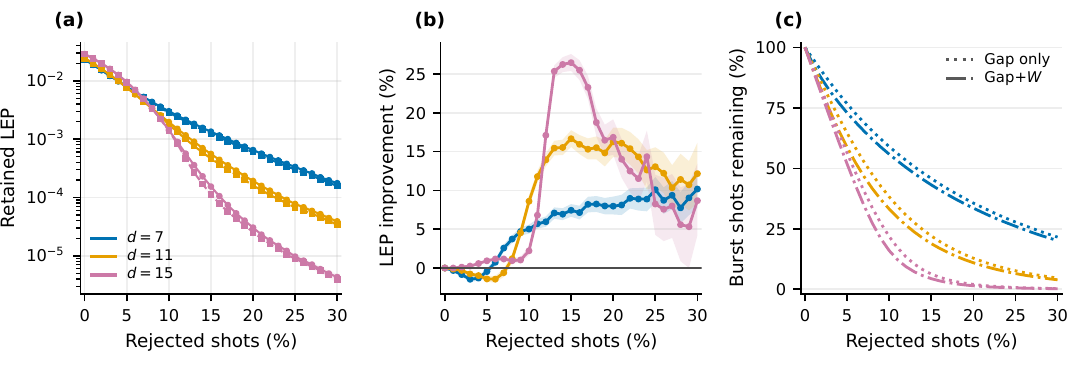}
  \caption{Circuit-level burst postselection with fixed $\alpha=0.05$, $r=d$, $\rho=0.1$, $p_{\mathrm{quiet}}=0.005$, and $p_{\mathrm{burst}}=0.010$ ($\xi=2$). (a) Retained LEP for gap-only and gap+$W$ selection at $d=7$, $11$, and $15$. Curves connect the one-percentage-point estimates; shaded regions are pointwise 95\% Wilson intervals. (b) Paired LEP improvement at matched rejection, with pointwise 95\% paired-bootstrap bands. (c) Fraction of the original burst-state population retained after gap-only (dotted) and gap+$W$ (dash-dotted) selection. Colors identify code distance throughout. The CDF transforms are fixed from independent stationary $p_0$ reference samples. The $d=7$, $11$, and $15$ curves use 8, 16, and 80 million shots, respectively.}
  \label{fig:circuit-burst-scaling}
\end{figure*}

Fig.~\ref{fig:gap-correction-residuals}(b) then connects this exact confidence miscalibration to the practical MWPM quantity $W$. Detector records are grouped at approximately fixed logical confidence and then resolved by their within-bin $W$ percentile. Using the MWPM gap for the grouping, high-$W$ records in the low-gap region acquire the largest positive risk response, indicating that records least compatible with the frozen decoder model are also those whose physical logical risk increases most rapidly under the mismatch. Repeating the same analysis using the exact reference gap $|\log[q_0(d)/q_1(d)]|$ produces the same qualitative dependence on $W$. The effect is therefore not simply correcting the ground-state approximation of the MWPM gap: even when the reference-model logical confidence is evaluated exactly, model mismatch leaves additional logical-risk information correlated with $W$.

The exact calculation therefore separates two effects that would otherwise be conflated. First, Fig.~\ref{fig:gap-correction-residuals}(a) shows that a logical gap that is exact for the frozen reference model can become miscalibrated when the physical noise distribution changes. Second, Fig.~\ref{fig:gap-correction-residuals}(b) shows that the MWPM cost $W$ identifies detector records for which this miscalibration is most detrimental, with the same qualitative behavior whether logical confidence is represented by the MWPM gap or by the exact reference gap. This provides the mechanism behind the likelihood-aware postselection behavior observed above.

\subsection{Scaling with code distance}

The exact calculation above is restricted to a single syndrome-extraction round and distance. We therefore test whether the same likelihood-aware postselection effect persists as both code distance and spacetime volume increase. Using the same Stim circuit-level noise model, we simulate rotated surface-code memories at $d=7,11,15$ with $r=d$ syndrome-extraction rounds. At each distance, the MWPM decoder is constructed from the stationary circuit at $p_0=0.005$ and then held fixed.

Physical shots use the same circuit-level noise channels with two whole-shot strengths. A fraction $1-\rho=0.9$ of shots use $p_{\mathrm{quiet}}=p_0=0.005$, while a fraction $\rho=0.1$ use
\begin{equation}
  p_{\mathrm{burst}}=\xi p_0.
  \label{eq:scaling-burst-noise}
\end{equation}
The same noise state is shared by every noisy circuit location within a shot, producing a fluctuation across the complete syndrome history. We fix $\xi=2$, so $p_{\mathrm{burst}}=0.010$ and the average location error probability is $(1-\rho)p_{\mathrm{quiet}}+\rho p_{\mathrm{burst}}=0.0055$. The postselection coefficient is $\alpha=0.05$ at every distance.

Figure~\ref{fig:circuit-burst-scaling} shows that likelihood-aware postselection continues to exploit bursts of noise as the code distance and spacetime volume increase. At matched rejection, gap+$W$ removes a larger fraction of burst-state shots than gap-only selection at every distance, with the separation becoming increasingly pronounced at larger $d$, as shown in Fig.~\ref{fig:circuit-burst-scaling}(c). The burst-state shots therefore become easier to preferentially reject as the code distance increases, producing the sharper intermediate-rejection improvement seen at larger distances. Once most burst-state shots have been removed, both selectors increasingly act on the quiet-state population and their performance converges. The coefficient $\alpha$ is held fixed across all distances and rejection fractions.

\section{Model mismatch on IBM hardware}

\begin{figure*}[t!]
  \centering
  \includegraphics[width=\textwidth]{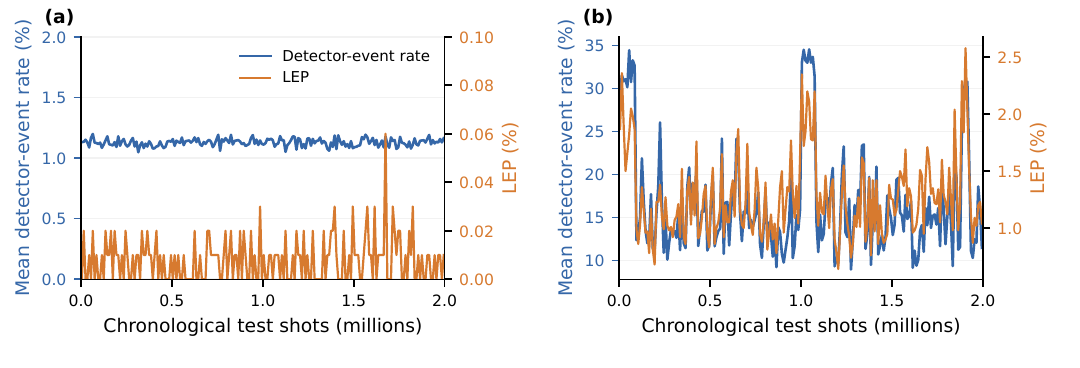}
  \caption{Two representative IBM repetition-code samples illustrating minimum and maximum detector-rate drift. (a) Distance-$5$ realization with the smallest detector-rate drift; (b) distance-$9$ realization with the largest drift. Blue shows the windowed detector-event rate and orange the hardware LEP obtained with the frozen calibration decoder. Each point represents a nonoverlapping 10,000-shot window. The examples are selected using detector statistics alone, without reference to logical outcomes; no confidence intervals are shown.}
  \label{fig:temporal-examples}
\end{figure*}
We now turn from controlled model mismatch to hardware data and ask whether the same ingredients arise during real quantum-memory experiments. Using chronologically ordered repetition-code measurements acquired on IBM hardware over approximately one month, we test whether the physical noise remains stationary over experimental timescales, whether periods of elevated syndrome activity persist and coincide with degraded logical performance, and whether the resulting detector histories remain consistent with the stationary DEM used for decoding. The IBM data provide the temporal resolution needed to expose these effects directly; later, we observe the same qualitative hardware--model mismatch and likelihood-aware postselection benefit in an independent Google surface-code memory dataset.

\subsection{Experimental records and frozen decoder}

The experiments were executed on the IBM \emph{Boston} and \emph{Kingston} processors through IBM Quantum Runtime~\cite{IBMQuantum}. Repetition-code circuits were generated in Stim~\cite{Gidney2021Stim} and translated and compiled through Qiskit~\cite{JavadiAbhari2024Qiskit} for execution on the physical backends. Jobs were submitted repeatedly over approximately one month, providing measurements at chronologically separated times rather than during a single acquisition. Each job contained multiple physical copies of the repetition code distributed across a processor, with layouts spanning the chip so that the dataset samples both different spatial regions and their evolution over time. Acquisition dates, physical-copy counts, code distances, and held-out shot counts are summarized in Appendix~\ref{app:experimental-data}.

The resulting dataset contains 163 repetition-code realizations at distances $d=5$, $7$, and $9$, with logical preparations in both the $Z$ and $X$ bases. Each realization contains chronologically ordered training and held-out records; the held-out half contributes approximately two million shots per realization and approximately 305 million held-out shots in total. We preserve acquisition order throughout the analysis and divide each held-out record into nonoverlapping 10,000-shot windows. Each shot prepares a known logical state, applies repeated syndrome extraction, and terminates in logical readout. The chronological training, validation, and evaluation data flow used throughout the hardware analysis is described in Appendix~\ref{app:common-coverage}.

For each physical realization, the calibration DEM is reconstructed from the IBM calibration snapshot associated with its acquisition and converted to a graphlike model for MWPM decoding. The resulting decoder is implemented with PyMatching~\cite{Higgott2022PyMatching,Higgott2025SparseBlossom} and frozen before the held-out record is analyzed, so temporal changes in the hardware are measured relative to a fixed noise model rather than absorbed by a time-dependent refit. The calibration DEM construction is given in Appendix~\ref{app:calibration-dem}. A second LEP-optimized decoder, which retains the same graphlike support but fits its mechanism rates using chronologically earlier training data, is introduced later for the postselection comparison; its optimization and validation procedure is given in Appendix~\ref{app:ler-optimized-decoder}.

Fig.~\ref{fig:temporal-examples} illustrates the range of temporal behavior observed during the acquisition campaign. The low-drift realization remains close to a detector-event rate of $1.1\%$, whereas the high-drift realization varies from approximately $8.9\%$ to $34.5\%$. In the latter record, the logical-error probability changes over the same sequence of windows, with periods of elevated detector activity accompanied by degraded logical performance. These changes persist across neighboring 10,000-shot windows rather than appearing only as isolated fluctuations, indicating transient high-noise conditions that extend over many consecutive shots.

\subsection{Temporal drift and logical degradation}

\begin{figure*}[t!]
  \centering
  \includegraphics[width=\textwidth]{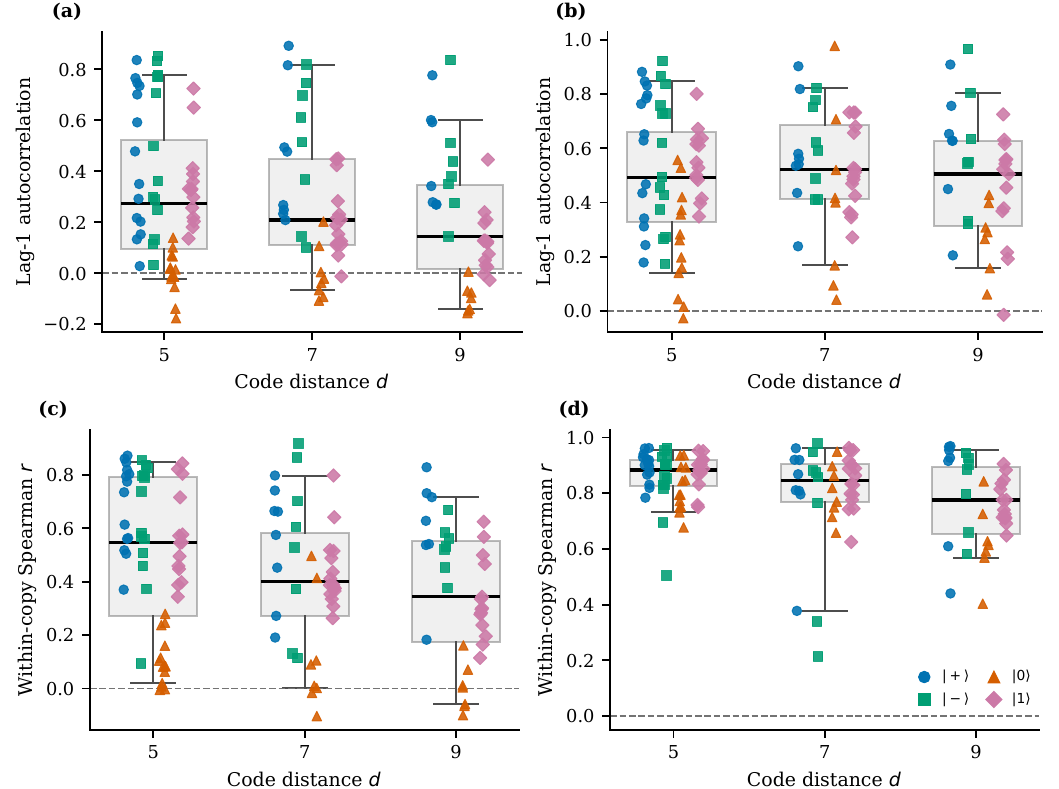}
  \caption{Temporal persistence and detector-rate association across IBM memory realizations. (a) Lag-one autocorrelation of LEP; (b) lag-one autocorrelation of the low-gap fraction; (c) Spearman correlation between detector-event rate and LEP, computed across chronological windows separately for each realization; (d) corresponding correlation between detector-event rate and low-gap fraction. Each point represents one physical realization; boxes show medians and interquartile ranges with 5th--95th percentile whiskers. These whiskers describe the empirical distribution across realizations and are not confidence intervals.}
  \label{fig:temporal-statistics}
\end{figure*}

We quantify this behavior across many physical layouts using both logical and syndrome-level observables. The detector-event rate measures the overall level of syndrome activity, while the low-gap fraction is defined as the fraction of shots whose MWPM gap lies in the lowest decile of that realization's validation distribution. The low-gap threshold is fixed before the held-out record is analyzed, so changes in this fraction reflect temporal changes in the distribution presented to the frozen decoder rather than adaptation of the threshold itself.

The temporal fluctuations are both persistent and associated with logical performance. Across the cohort, the median lag-one autocorrelation is $0.22$ for LEP and $0.50$ for the low-gap fraction, showing that elevated-error conditions extend across neighboring 10,000-shot windows. Computing the Spearman correlation across chronological windows separately for each realization, the median correlation between detector-event rate and LEP is $0.46$, while the corresponding median correlation with the low-gap fraction is $0.85$. Periods of elevated syndrome activity are therefore accompanied by increased logical-error probabilities and by a larger concentration of detector records that the frozen decoder regards as logically ambiguous.

The stronger association with the low-gap fraction is consistent with its lower sampling noise, since every shot contributes a decoder gap whereas logical failures can remain sparse within an individual 10,000-shot window. Taken together, the temporal autocorrelations and per-realization correlations show that the processor undergoes transient noise whose effects persist over experimentally relevant timescales and are accompanied by coherent degradation of memory performance.

\subsection{High-cost tails reveal model mismatch}

\begin{figure*}[t]
  \centering
  \includegraphics[width=\textwidth,trim=8pt 8pt 8pt 8pt,clip]{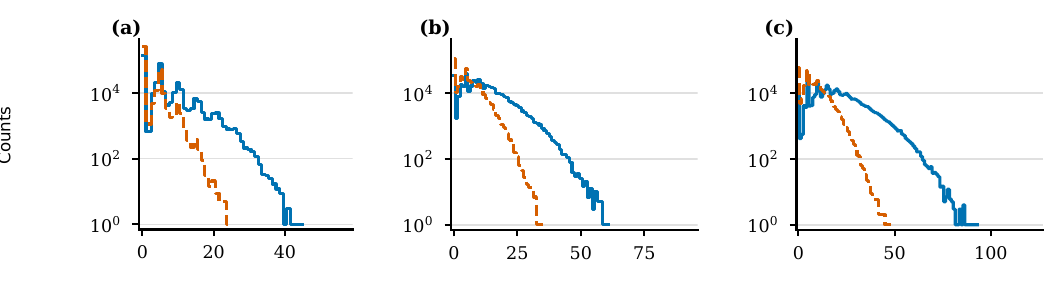}
  \caption{Hardware and calibration-DEM distributions of the minimum MWPM cost $W$ for the logical $|1\rangle$ state. (a) $d=5$; (b) $d=7$; (c) $d=9$. Solid blue curves show QPU data and dashed orange curves show calibration-DEM samples; counts are logarithmic. The other preparations are shown in Appendix~\ref{app:weight-mismatch-all-states}. No uncertainty intervals are shown for these pooled counts.}
  \label{fig:weight-mismatch}
\end{figure*}

The temporal analysis establishes that the physical conditions encountered by a frozen decoder vary substantially during the experiment. We next test whether the stationary calibration DEM reproduces the detector histories generated under these conditions by comparing the minimum MWPM cost $W$ on held-out hardware data with the distribution obtained from independent records sampled from the corresponding calibration DEM. Because $W$ is the minimum cost assigned to a detector record by the decoder model, an excess of high-$W$ events indicates detector histories that are substantially less compatible with the assumed model than its own samples predict.

For each realization $j$, we quantify the displacement of the high-cost tail using the 99th-percentile cost ratio
\begin{equation}
  \kappa_{99}^{(j)}
  =
  \frac{
  q_{0.99}\!\left(W_{\mathrm{QPU}}^{(j)}\right)
  }{
  q_{0.99}\!\left(W_{\mathrm{DEM}}^{(j)}\right)
  },
  \label{eq:weight-tail-quantile-ratio}
\end{equation}
where $q_{0.99}$ denotes the sample 99th percentile. Across all 163 realizations, the median value is $\kappa_{99}=2.30$, and $25.7\%$ of held-out hardware shots lie above the corresponding DEM 99th-percentile threshold. The detailed DEM-sampling procedure and finite-sample construction of this comparison are given in Appendix~\ref{app:dem-sampling-weight}.

The heavy hardware tail in Fig.~\ref{fig:weight-mismatch} shows that the calibration DEM substantially under-represents detector histories that occur on the processor. The complete state-resolved comparison is given in Appendix~\ref{app:weight-mismatch-all-states}. Together with the chronological analysis, this demonstrates that the physical noise sampled during the experiment can depart strongly from the stationary model assumed by the decoder, despite that decoder remaining useful as a logical decision rule. The month-long acquisition and the use of multiple physical code realizations spanning the processor further show that this discrepancy is not confined to a single short run or isolated region of the device.

In the next section, we exploit this hardware--model mismatch using likelihood-aware gap postselection. By supplementing the logical gap with the absolute model-compatibility information contained in $W$, we identify and preferentially reject detector records that are poorly described by the frozen decoder model, reducing the retained logical-error probability on held-out hardware data.

\section{Likelihood-aware postselection on hardware}

The preceding section established that detector records produced on hardware can depart substantially from the stationary model used by a frozen decoder. We now ask whether that mismatch can be exploited operationally. We apply likelihood-aware postselection to independent IBM repetition-code and Google surface-code memory experiments, using the minimum MWPM cost $W$ to supplement the logical gap while keeping the retained-shot count exactly matched to gap-only postselection. All likelihood coordinates, postselection coefficients, and operating thresholds are determined from chronologically earlier development and validation data and frozen before held-out evaluation. The two datasets provide complementary tests across different processors, code families, decoder priors, state-preparation protocols, and experimental timescales.

\subsection{IBM repetition-code memories}

We first evaluate likelihood-aware postselection on the held-out IBM repetition-code records introduced in the previous section. The analysis contains 163 physical realizations at distances $d=5$, $7$, and $9$, with approximately two million held-out shots per realization. For every realization and target coverage, the likelihood-aware coefficient $\alpha$ is selected using only the chronological development and validation records and then frozen before the second chronological half is evaluated. Gap-only and likelihood-aware postselection are compared at exactly matched retained-shot counts; the complete data flow and common-coverage procedure are given in Appendix~\ref{app:common-coverage}.

We evaluate two frozen decoder models to distinguish the postselection effect from the quality of the underlying decoder prior. The calibration decoder is constructed directly from the IBM calibration snapshot associated with each physical realization. The LEP-optimized decoder retains the same graphlike detector and logical support but adapts its mechanism rates using only earlier chronological records. Its training proceeds in two stages: the rates are first adjusted to reproduce projected training-syndrome distributions and are then further optimized for full-coverage logical-error probability on development data, with validation data selecting the retained parameter set. No held-out shot enters either optimization stage. The full construction is described in Appendices~\ref{app:calibration-dem} and~\ref{app:ler-optimized-decoder}.

\begin{figure*}[t]
  \centering
  \includegraphics[width=\textwidth]{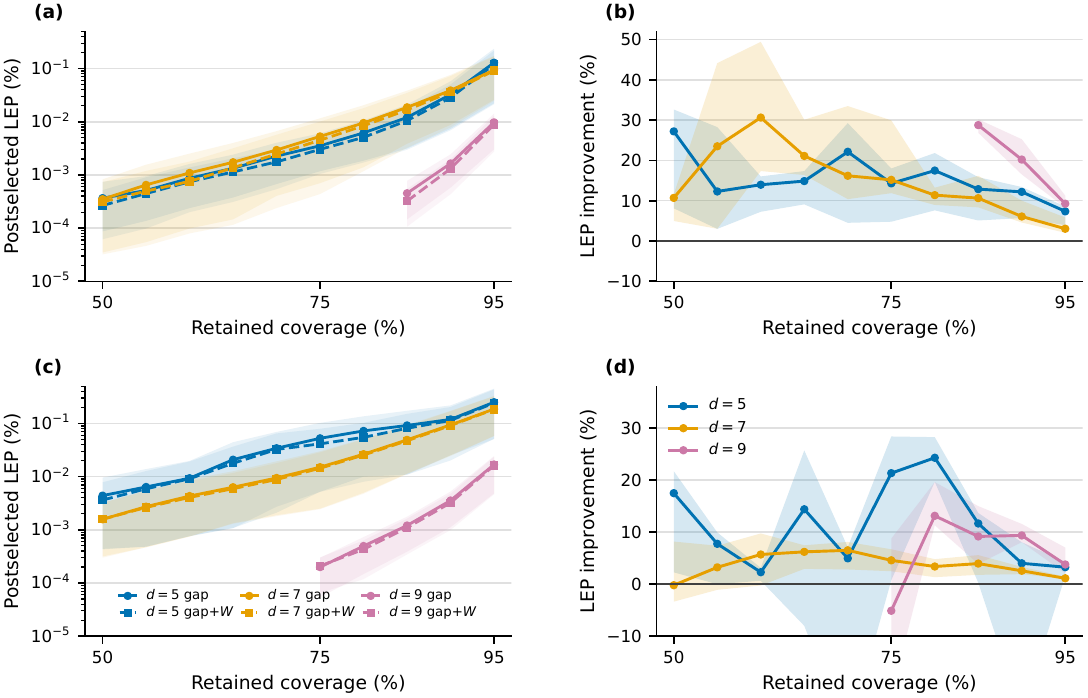}
  \caption{IBM retained-coverage curves for gap-only and likelihood-aware postselection. (a) Held-out LEP with the LEP-optimized decoder; (b) corresponding relative LEP reduction; (c) held-out LEP with the calibration DEM; (d) corresponding relative reduction. Curves are separated by distance and span 50--95\% retained coverage. Bands are pointwise 95\% percentile intervals from 20,000 paired acquisition-job bootstrap replicates, with score selection and coverage matching fixed. Score selection and coverage matching are described in Appendix~\ref{app:hardware-methods}.}
  \label{fig:ibm-improvement}
\end{figure*}

The distance-resolved results show a substantial reduction in retained logical errors. At 15\% rejection, likelihood-aware postselection reduces the held-out LEP relative to gap-only postselection by $10.64\%$ with a 95\% interval of $[8.35\%,16.18\%]$ at $d=7$ and by $28.75\%$ $[26.84\%,30.20\%]$ at $d=9$. Fig.~\ref{fig:ibm-improvement} shows the retained-coverage dependence for both decoder models, while the fixed-rejection comparison is given in Appendix~\ref{app:supplemental-hardware}. The improvement across multiple distances and physical code realizations shows that the model-compatibility coordinate can refine the logical gap on held-out hardware records rather than only in the controlled mismatch constructions considered in simulation.

The comparison between decoder models is also informative. Likelihood-aware postselection remains beneficial after the graphlike prior has been explicitly optimized for full-coverage hardware LEP, showing that the effect is not restricted to the original calibration DEM. At the same time, changing the decoder prior changes both $\Delta$ and $W$ and therefore changes the residual information available to the postselection score. Decoder optimization and likelihood-aware postselection should consequently be viewed as complementary operations: the former improves the model used to make the logical decision, while the latter uses residual model incompatibility to refine the confidence assigned to that decision.

\subsection{Google surface-code memories}

We next test the same construction on the public Google Quantum AI surface-code data associated with the below-threshold memory experiments~\cite{GoogleBelowThreshold2025,GoogleDataset2024}. The measurements were acquired on a 72-qubit Willow superconducting processor and contain distance-$d=3$ and $d=5$ surface-code memories in the logical $X_L$ and $Z_L$ bases over a range of syndrome-extraction durations~\cite{GoogleBelowThreshold2025}. This changes the physical processor, stabilizer geometry, circuit construction, and noise environment relative to the IBM repetition-code experiment and therefore provides an independent test of likelihood-aware postselection.

The Google state-preparation protocol differs from the fixed logical-state preparations used in the IBM experiments. For each logical basis, the data qubits are initialized in physical product states corresponding to eigenstates of either $X_L$ or $Z_L$~\cite{GoogleBelowThreshold2025}. Within each code, basis, and circuit duration, the experiment samples multiple randomized initialization bitstrings and their complements, so the released data contain both logical eigenvalues and multiple physical representatives within a given logical basis~\cite{GoogleBelowThreshold2025,GoogleDataset2024}. We consequently report the Google results by logical basis rather than by the individually prepared logical states $|0_L\rangle$, $|1_L\rangle$, $|+_L\rangle$, and $|-_L\rangle$ used in the IBM analysis.

We evaluate two frozen matching priors. The first is the RL-optimized prior, obtained using the decoder-prior optimization introduced by Sivak \emph{et al.}, which adjusts matching priors to minimize hardware logical-error probability~\cite{Sivak2024}. The second is a prior derived from the device-independent SI1000 circuit-level Pauli-noise model~\cite{Gidney2021Honeycomb}. We do not refit either prior using the held-out records. Earlier acquisition samples define the likelihood coordinates and select $\alpha$, while eight later acquisition samples are reserved for held-out evaluation, as summarized in Appendix~\ref{app:common-coverage}. The chronological sample ordering and dataset conventions follow the released Google Quantum AI records~\cite{GoogleDataset2024}.

\begin{figure*}[t]
  \centering
  \includegraphics[width=\textwidth]{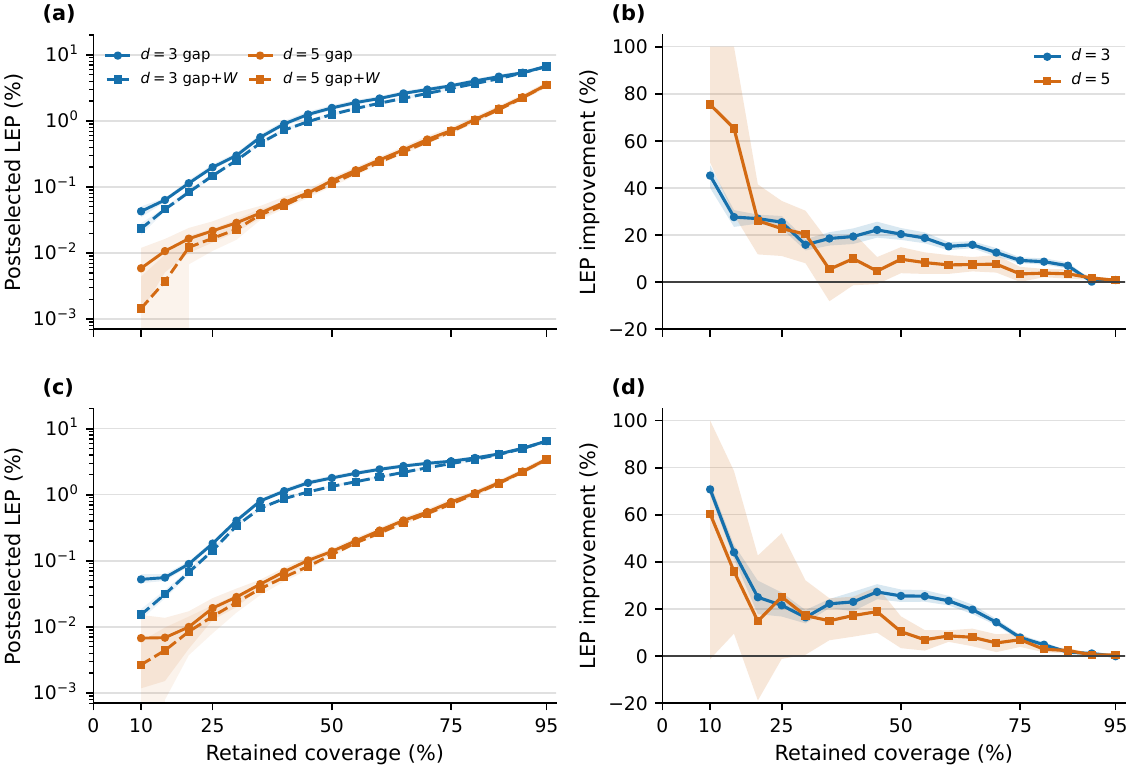}
  \caption{Google retained-coverage curves at ten syndrome-extraction rounds for both supplied decoder priors. (a) Held-out postselected LEP for the RL-optimized prior; (b) corresponding relative LEP reduction; (c) held-out LEP for the SI1000 prior; (d) corresponding relative reduction. Every panel includes both $d=3$ and $d=5$. LEP is logarithmic, and shaded regions are pointwise 95\% percentile intervals from 20,000 paired bootstrap replicates over the eight held-out acquisition samples, with score choices fixed.}
  \label{fig:google-retained-coverage}
\end{figure*}

The Google data reproduce the likelihood-aware advantage in a distinct surface-code setting. For the ten-round $d=5$ memory experiment, likelihood-aware postselection reduces the retained LEP by $75.4\%$ with a 95\% interval of $[50.8\%,100\%]$ at 90\% rejection relative to gap-only postselection. Fig.~\ref{fig:google-retained-coverage} shows the retained-coverage comparison at ten rounds for both priors. This duration provides a representative operating point with improvement across a broad rejection range; the round-resolved comparison and its conditional-information diagnostic appear together in Appendix~\ref{app:supplemental-hardware}. The size of the effect varies with distance, operating coverage, and decoder prior, consistent with the expectation that the information remaining in $W$ depends on how the frozen decoder represents the physical noise.

Across both hardware platforms, the data also provide suggestive evidence that priors explicitly optimized for logical performance are more robust to the underlying hardware variation. The IBM LEP-optimized decoder and Google's RL-optimized prior~\cite{Sivak2024} both remain effective under likelihood-aware postselection and appear less sensitive to model mismatch in portions of the retained-coverage and temporal data. We do not attempt to quantify this apparent stability or determine which features of the optimization are responsible for it. Jointly studying decoder-prior optimization, temporal robustness, and likelihood-aware confidence under drift is therefore an important direction for future work.

Taken together, the IBM and Google memory experiments show that likelihood-aware postselection can improve logical-gap postselection across repetition- and surface-code memories, independently acquired superconducting processors, and both calibration-derived and hardware-optimized decoder priors. We next test whether the same model-compatibility principle extends beyond quantum memory to the distinct task of magic-state cultivation.

\subsection{Magic-state cultivation}

\begin{figure*}[t]
    \centering
    \includegraphics[width=\textwidth]{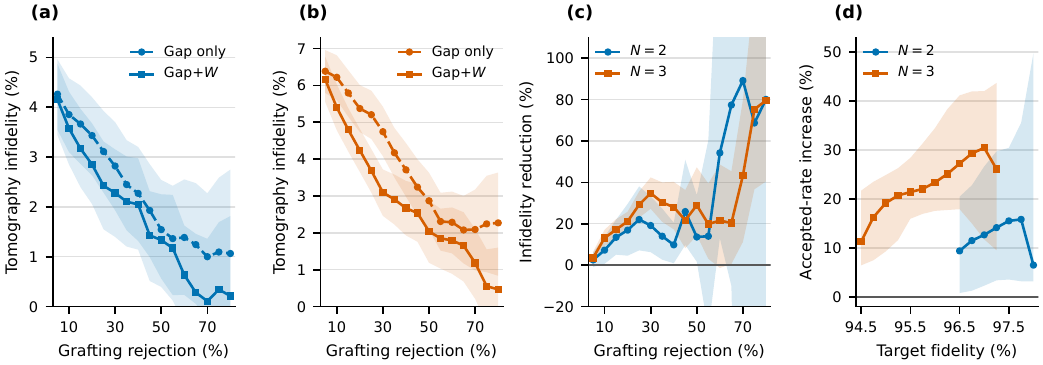}
    \caption{Likelihood-aware postselection applied to the grafting stage of the Google magic-state cultivation experiment. Google's native cultivation acceptance is applied before the additional postselection considered here. (a) Absolute tomography infidelity versus additional grafting-stage rejection for two-cycle cultivation; (b) corresponding results for three-cycle cultivation. Dashed curves show gap-only postselection and solid curves show likelihood-aware postselection. (c) Relative tomography-infidelity reduction at fixed retained rate as additional grafting-stage rejection increases. (d) Relative accepted-rate increase at fixed target tomography fidelity. Shaded regions are pointwise 95\% percentile intervals from paired 500-shot acquisition-order block bootstraps, using 2,000 replicates for (c) and 4,000 for (a), (b), and (d), as detailed in Appendix~\ref{app:cultivation-methods}.}
    \label{fig:google-cultivation}
\end{figure*}

Fault-tolerant quantum computation requires non-Clifford operations in addition to the protected Clifford operations naturally available in many error-correcting architectures. Magic states provide these operations through state injection and are therefore a central resource for universal fault-tolerant computation~\cite{BravyiKitaev2005}. Their preparation can also constitute a substantial fraction of the spacetime cost of a fault-tolerant architecture, motivating both optimized magic-state factories and alternatives to conventional distillation~\cite{Litinski2019,Gidney2024}. Magic-state cultivation is one such alternative: rather than combining many encoded noisy states through repeated logical distillation, it progressively increases the reliability of a single encoded $|T\rangle$ state using fault-tolerant checks, code growth, and postselection~\cite{Gidney2024,Bombin2024}.

The cultivation construction is naturally divided into injection, cultivation, and escape stages~\cite{Gidney2024}. Injection creates an initial encoded magic state in a small code. Cultivation then applies logical cross-checks and stabilizer measurements, rejecting attempts when the measurement record indicates a fault and thereby progressively increasing the fault distance of the surviving state. The resulting state can become substantially more reliable than the small code that contains it, so it must subsequently escape into a larger encoding before ordinary unpostselected error-correction cycles erase the fidelity gained during cultivation. Gidney \emph{et al.} introduced grafting as a rapid route for this escape and emphasized that the escape stage cannot efficiently reject every nontrivial detector record, making graded decoder confidence particularly relevant in this part of the protocol~\cite{Gidney2024}.

We analyze the released Google experimental realization of this protocol~\cite{Rosenfeld2025,GoogleCultivationData2025}. The experiment performs injection and cultivation in a distance-three color code and then grafts the surviving magic state into a larger distance-five encoding compatible with subsequent surface-code operation~\cite{Rosenfeld2025}. The intermediate grafted code contains both color-code-like and surface-code stabilizers and is not directly matchable; Google therefore decodes the grafting record using Tesseract, an $A^*$-based most-likely-error decoder~\cite{Rosenfeld2025,Tesseract2025}.

We specifically target the grafting stage. The hard cultivation acceptance is applied first and is left unchanged, so every record considered here has already passed the fault-detection checks used to accept the cultivated state. We then project the pre-tomography Tesseract decoding problem into the two relevant logical sectors and obtain the complementary gap $\Delta$ and minimum sector cost $W$ from their constrained costs. The same likelihood-aware construction used above supplements the relative logical confidence carried by $\Delta$ with the model-compatibility information carried by $W$. A single basis-blind score is selected from chronologically earlier records without using held-out tomography outcomes and is frozen before evaluation. The ordered split, score construction, and paired block-bootstrap intervals are specified in Appendices~\ref{app:common-coverage} and~\ref{app:cultivation-methods}.

This experiment also makes the acceptance--quality tradeoff operationally important. Cultivation is intrinsically retryable: rejecting an additional candidate can improve the fidelity of the states that survive, but it also consumes another preparation attempt before a usable resource is delivered~\cite{Gidney2024,Bombin2024}. In a fault-tolerant architecture supplied by magic-state factories, the relevant resource is therefore not fidelity alone but the rate at which states satisfying the required fidelity can be produced~\cite{Litinski2019}. We consequently report likelihood-aware postselection in two complementary forms: the reduction in tomography infidelity at fixed additional rejection and the increase in accepted rate at fixed target fidelity. The latter provides a direct measure of the potential throughput benefit when magic-state production is acceptance limited, although translating it into an end-to-end computational speedup would additionally require modeling factory parallelism, buffering, and downstream consumption.

We observe in Fig.~\ref{fig:google-cultivation} that additional postselection during the grafting stage further improves the quality of the final state, with additional gains from the likelihood-aware technique. For the three-cycle data, likelihood-aware postselection reduces held-out tomography infidelity by $34.74\%$ relative to gap-only postselection at 30\% additional rejection, with a 95\% paired block-bootstrap interval of $[27.21\%,42.39\%]$. The two-cycle data show the same effect, with a $19.15\%$ reduction and interval $[6.28\%,33.22\%]$ at the same operating point. Fig.~\ref{fig:google-cultivation}(c) shows that the improvement persists over a broader range of additional rejection before the uncertainty increases in the sparsest retained tail.

The same advantage can instead be used to increase yield while holding the required output quality fixed. For the three-cycle data, likelihood-aware postselection increases the accepted rate by $30.44\%$ $[11.73\%,42.04\%]$ at a target tomography fidelity of $97\%$. For the two-cycle data, the accepted rate increases by $15.51\%$ $[3.52\%,30.44\%]$ at a target fidelity of $97.5\%$. At fixed fidelity, these states satisfy the same output-quality requirement while requiring fewer discarded preparations, so an otherwise acceptance-limited resource-state pipeline could supply usable magic states at a higher rate.

The cultivation experiment extends the likelihood-aware principle beyond quantum memory in two important respects. First, the added score operates after an existing fault-tolerant postselection protocol rather than replacing its native checks, showing that soft decoder information can further rank preparations that have already passed a hard acceptance criterion. Second, the construction transfers from MWPM to the Tesseract most-likely-error decoder by using the same underlying pair of quantities: relative preference between logical sectors and absolute cost of the preferred explanation. Together with the IBM and Google memory results, this suggests that likelihood-aware postselection can act as a lightweight control layer for the quality--throughput tradeoff of future fault-tolerant resource factories. Determining how much of the accepted-rate improvement translates into end-to-end computational throughput under real-time decoding, buffering, parallel factory operation, and magic-state consumption remains an important direction for future work.

\section{Discussion and outlook}

Fault-tolerant quantum computation relies on decoders not only to infer corrections from syndrome measurements, but also, increasingly, to provide soft information about the reliability of those decisions. Such decoder confidence can be interpreted as a record-dependent estimate of logical-error risk and can support downstream tasks that depend on the reliability of a logical outcome. Postselection is one important example, where confidence is used to trade acceptance rate for improved logical fidelity. The usefulness of such soft information, however, depends on how accurately the reported confidence reflects the physical probability of logical failure. For an exact decoder supplied with the true physical noise distribution, the logical gap completely determines the conditional logical-error probability. When the physical distribution differs from the model assumed by the decoder, this correspondence can break down and the reported confidence can become miscalibrated.

The central distinction exposed by this work is is between relative logical confidence and compatibility with the decoder model. The logical gap measures how strongly the assumed model favors one logical sector over another, whereas model compatibility measures how well the observed detector record is supported by that model. Our analysis shows that compatibility can provide additional information when the physical perturbation reweights the competing logical sectors differently: a detector record may become much less likely without becoming more logically dangerous if both sectors are affected in the same way. Model compatibility therefore does not by itself determine logical risk, but it can help identify records for which the confidence reported by a frozen decoder is miscalibrated.

The hardware and simulation results indicate that this effect can arise in practice. On IBM hardware, syndrome statistics vary persistently in time and the observed detector histories contain a substantially heavier high-cost tail than predicted by the frozen decoder model. A related discrepancy between observed detector histories and the assumed decoder model is also present in the independent Google surface-code memory data. Controlled burst simulations illustrate how latent shot-to-shot variation can generate correlations outside a stationary independent detector error model, while the exactly enumerable circuit shows that, in this setting, even the exact logical gap of a fixed reference model becomes miscalibrated as the physical distribution changes. Simulations at larger code distance and spacetime volume further indicate that burst-state records can become increasingly distinguishable through their accumulated decoder cost, so this source of model mismatch can remain relevant as the encoded system grows.

This additional information leads to measurable improvements in postselection across the hardware datasets considered here. At $15\%$ rejection, likelihood-aware postselection reduces the retained logical-error probability by $28.8\%$ on the held-out IBM $d=9$ repetition-code memories. On ten-round Google $d=5$ XZZX surface-code memories, the reduction is $75.4\%$ at $90\%$ rejection. The same construction also transfers across changes in processor, code family, decoder prior, and experimental protocol. In the cultivation experiment, where the method is applied after Google's native cultivation checks, the additional decoder information reduces tomography infidelity by $34.7\%$ at the reported operating point, or can instead be used to increase the accepted yield by $30.4\%$ at fixed target fidelity.

The cultivation result provides one example of how this information could be used in a fault-tolerant resource-preparation setting. Resource-state factories must balance the fidelity of accepted states against the rate at which those states are produced. Hard fault-tolerance checks provide one stage of this selection, while decoder soft information can further rank states that have already passed those checks. In this setting, likelihood-aware postselection changes the quality--throughput tradeoff rather than only the logical error rate. Decoder-model compatibility could therefore provide an additional signal for deciding which prepared resources to accept or discard before they are consumed by a larger fault-tolerant computation.

Although we have used MWPM and the Tesseract decoder in the experiments studied here, the underlying model-mismatch issue is not specific to either decoder. In principle, any decoder that assigns confidence using an assumed model of the physical noise can become miscalibrated when the true distribution contains correlations, temporal variation, latent structure, or other effects absent from that model. It would therefore be useful to identify analogous model-compatibility quantities for other decoder families. For MWPM, the minimum cost $W$ is convenient because it is inexpensive and already available from the sector-constrained decoding problem, but it should be viewed as one possible proxy rather than an optimal measure. Other decoders may provide likelihoods, local scores, latent-state estimates, or learned features that serve a similar role or more directly identify where their reported confidence becomes unreliable.

More generally, these results suggest that decoder confidence can benefit from separating two questions: which logical outcome is preferred under the assumed model, and how well that model describes the detector record being decoded. The logical gap addresses the first question, while model-compatibility information provides partial information about the second. Incorporating both quantities may provide a useful way to refine postselection when the physical noise departs from the model used by the decoder, including in larger codes and fault-tolerant resource-preparation protocols.
\begin{acknowledgments}
This material is based upon work supported by the U.S. Department of Energy, Office of Science, National Quantum Information Science Research Centers, Quantum Science Center (QSC). This research was supported by PNNL's Quantum Algorithms and Architecture for Domain Science (QuAADS) Laboratory Directed Research and Development (LDRD) Initiative. The Pacific Northwest National Laboratory is operated by Battelle for the U.S. Department of Energy under Contract No. DE-AC05-76RL01830. This research used resources of the Oak Ridge Leadership Computing Facility (OLCF), a DOE Office of Science User Facility supported under Contract No. DE-AC05-00OR22725. This research used resources of the National Energy Research Scientific Computing Center (NERSC), a U.S. Department of Energy Office of Science User Facility located at Lawrence Berkeley National Laboratory and operated under Contract No. DE-AC02-05CH11231, using NERSC awards DDR-ERCAP0038362, ASCR-ERCAP0037552, and DDR-ERCAP0038957.

OpenAI ChatGPT, using GPT-6 Sol, was used as a learning and research-assistance tool, including assistance with the development and debugging of numerical simulations and drafting and improving the clarity of the manuscript. All AI-assisted code, figures, scientific content, and text were independently reviewed, tested, and verified by the authors, who take full responsibility for the final manuscript and reported results.

\end{acknowledgments}
\appendix


\section{Datasets and analysis protocol}
\label{app:experimental-data}

This appendix collects the experimental data taken and frozen data splits used throughout the hardware analyses. The three datasets serve different purposes: the IBM repetition-code records resolve temporal variation across repeated hardware acquisitions, the Google surface-code records test likelihood-aware postselection on an independent memory experiment and decoder family, and the Google cultivation records test the same principle in a retryable fault-tolerant state-preparation protocol. In every case, score coordinates and hyperparameters are determined before the held-out records used for the reported performance estimates are accessed.

\subsection{IBM repetition-code dataset}
\label{app:ibm-data}

The primary IBM dataset contains 163 physical repetition-code realizations acquired on the \emph{Boston} and \emph{Kingston} processors through IBM Quantum Runtime~\cite{IBMQuantum}. Circuits were generated in Stim~\cite{Gidney2021Stim}, translated and compiled through Qiskit~\cite{JavadiAbhari2024Qiskit}, and executed repeatedly over the acquisition campaign. The dataset contains distances $d=5$, $7$, and $9$ and logical preparations in both the $X$ and $Z$ bases. Each acquisition job contains multiple physical copies distributed across the processor, so the complete dataset samples both different device regions and different acquisition times.

Each physical realization is divided chronologically into training and held-out portions before decoder optimization or postselection tuning. The primary cohort used for the IBM retained-coverage analysis contains approximately $305$ million held-out shots. We preserve acquisition-job and physical-copy identities throughout the analysis so that statistical resampling reflects hardware-level variation rather than treating individual shots as independent experimental replicates.

The source table below records the dataset identifier, acquisition job, code distance, prepared logical state, number of physical copies, and chronological held-out shot count used in the analysis.

\subsection{Google surface-code memory dataset}
\label{app:google-memory-data}

The Google memory analysis uses the public \texttt{google\_72Q\_surface\_code\_d3\_d5\_set1} release associated with the below-threshold surface-code experiments~\cite{GoogleBelowThreshold2025,GoogleDataset2024}. The released circuits use the XZZX convention and contain five distance-$3$ layouts, \texttt{d3\_at\_q3\_5}, \texttt{d3\_at\_q5\_3}, \texttt{d3\_at\_q5\_5}, \texttt{d3\_at\_q5\_7}, and \texttt{d3\_at\_q7\_5}, together with the distance-$5$ layout \texttt{d5\_at\_q5\_5}. We analyze logical $X$- and $Z$-basis memories at $r=1,10,13,30,50,70,90,110,130,150,170,190,210,230,$ and $250$ syndrome-extraction rounds.

\begin{center}
  \captionof{table}{IBM repetition-code data used in this work. Each experiment uses a number of QEC rounds equal to its code distance, $T=d$. Shot counts are chronological held-out hardware shots summed over the listed physical copies; the complete dataset identifiers are in \texttt{source\_data/table\_ibm\_data.csv}.}
  \label{tab:ibm-data-appendix}
  \setlength{\tabcolsep}{3pt}
  \resizebox{0.97\linewidth}{!}{%
      \begin{tabular}{lllrrrr}
        \toprule
        Backend & Acquisition & State & $d$ & Rounds & Copies & Test shots \\
        \midrule
        Boston & 2026-08-10 & $|+\rangle$ & 5 & 5 & 14 & 28,000,000 \\
        Boston & 2026-08-10 & $|-\rangle$ & 5 & 5 & 14 & 28,000,000 \\
        Boston & 2026-08-08 & $|0\rangle$ & 5 & 5 & 14 & 28,000,000 \\
        Boston & 2026-08-08 & $|1\rangle$ & 5 & 5 & 14 & 28,000,000 \\
        Boston & 2026-08-11 & $|+\rangle$ & 7 & 7 & 8 & 16,000,000 \\
        Boston & 2026-08-11 & $|-\rangle$ & 7 & 7 & 8 & 16,000,000 \\
        Boston & 2026-07-29 & $|0\rangle$ & 7 & 7 & 8 & 16,000,000 \\
        Boston & 2026-07-17 & $|1\rangle$ & 7 & 7 & 8 & 16,000,000 \\
        Boston & 2026-07-20 & $|1\rangle$ & 7 & 7 & 8 & 16,000,000 \\
        Boston & 2026-09-11 & $|+\rangle$ & 7 & 7 & 8 & 16,000,000 \\
        Boston & 2026-09-11 & $|-\rangle$ & 7 & 7 & 8 & 16,000,000 \\
        Boston & 2026-09-11 & $|0\rangle$ & 7 & 7 & 8 & 16,000,000 \\
        Boston & 2026-09-11 & $|1\rangle$ & 7 & 7 & 8 & 16,000,000 \\
        Boston & 2026-08-11 & $|+\rangle$ & 9 & 9 & 7 & 14,000,000 \\
        Boston & 2026-08-11 & $|-\rangle$ & 9 & 9 & 7 & 14,000,000 \\
        Boston & 2026-08-03 & $|0\rangle$ & 9 & 9 & 7 & 7,000,000 \\
        Boston & 2026-08-03 & $|1\rangle$ & 9 & 9 & 7 & 7,000,000 \\
        Kingston & 2026-08-03 & $|1\rangle$ & 9 & 9 & 7 & 7,000,000 \\
        \bottomrule
      \end{tabular}
  }
\end{center}

Each layout, basis, and round-count experiment contains ten initialization patterns with $5{,}000$ shots per pattern, giving $50{,}000$ shots per acquisition sample. The ten patterns form five bitwise-complement pairs. Because the protected logical string has odd weight, global complementation exchanges the two logical eigenvalues; the released logical-observable label is defined relative to the noiseless result for each initialization pattern. We therefore pool the randomized physical representatives within each logical basis rather than fitting a separate postselection rule for each initialization bitstring.

Samples 05--08 are used for development, samples 09--12 for validation, and samples 13--20 for held-out evaluation. The analysis uses detector records and the released logical-observable outcomes; the initialization sweep bits are not used as postselection features. This separation allows the score coordinates and coefficient $\alpha$ to be fixed using earlier acquisition samples before the eight later samples are evaluated.

\begin{figure}[t]
  \centering
  \includegraphics[width=\columnwidth]{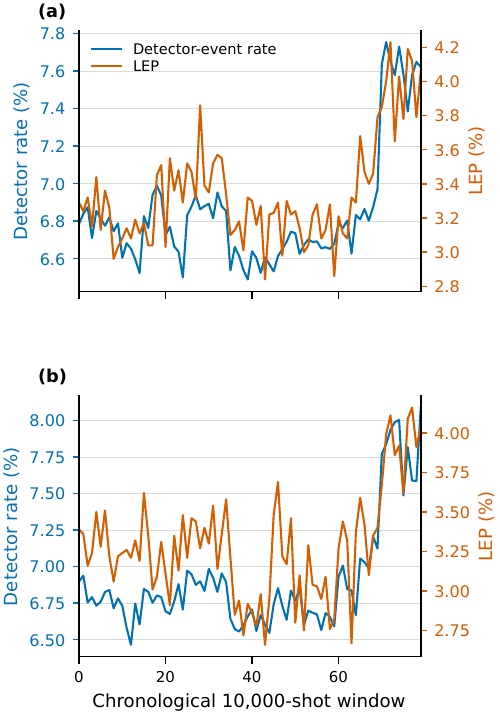}
  \caption{Chronological detector activity and frozen-decoder logical performance in the Google surface-code memory data. The two panels use the centered $d=5$, ten-round layout in the logical $X$ and $Z$ bases, respectively, and concatenate the chronological acquisition samples 05--20. Each point is a nonoverlapping 10,000-shot window. Blue gives the detector-event rate and orange gives the logical-error probability obtained from the supplied RL-optimized decoder output. This diagnostic is descriptive; the postselection coordinates and operating points are still fixed using the earlier development and validation samples. No uncertainty intervals are shown.}
  \label{fig:google-temporal-drift}
\end{figure}

Additionally in \ref{fig:google-temporal-drift} we also observe temporal variation of the detector distribution across multiple samples. Which also leads to a time varying LEP. 

\subsection{Google magic-state cultivation dataset}
\label{app:cultivation-data}

The cultivation analysis uses the released Google two-cycle and three-cycle magic-state cultivation records~\cite{Rosenfeld2025,GoogleCultivationData2025}. Google's native cultivation acceptance is applied before any likelihood-aware analysis. In particular, the first 16 state-preparation detectors are required to be zero; records that pass this hard criterion remain eligible even when later grafting or error-correction detectors fire.

For two-cycle cultivation, the pre-tomography record contains detectors $0$--$52$, giving 53 detectors in total, while the three-cycle record contains detectors $0$--$79$, giving 80 detectors. All detector information after the pre-tomography prefix is excluded from the postselection score. The resulting eligible records are ordered according to the released row order separately within each measurement basis and cultivation depth.

The held-out tomography analysis combines $X$- and $Z$-basis measurements to estimate the target-state fidelity. Postselection features are constructed only from the pre-tomography grafting record; tomography outcomes are used for development and validation objectives where required and for final held-out evaluation, but never enter the feature vector applied to an individual held-out shot.

\subsection{Ordered development, validation, and test splits}
\label{app:common-coverage}

The common design principle is ordered separation between model construction and evaluation. IBM decoders and postselection coordinates are trained on the first chronological half of each physical realization and evaluated on the second half. Google memory coordinates are fit on samples 05--08, $\alpha$ is selected on samples 09--12, and samples 13--20 are held out. The cultivation release does not provide one global acquisition chronology across tomography bases. We therefore preserve the released row order separately within each basis and cultivation depth: the first quarter constructs the coordinates, the second quarter selects $\alpha$, the coordinates are refit on the first half, and the final half is held out.

\begin{table}[t]
\caption{Frozen data flow for the three hardware analyses. ``Coordinates'' denotes the data used to construct empirical score transforms; no held-out logical or tomography outcome is used to choose $\alpha$ or a postselection threshold.}
\label{tab:hardware-data-flow}
\centering
\scriptsize
\begin{tabular}{@{}p{0.16\columnwidth}p{0.28\columnwidth}p{0.22\columnwidth}p{0.24\columnwidth}@{}}
\toprule
Data & Coordinates and model selection & Held-out evaluation & Uncertainty unit \\
\midrule
IBM memory & First chronological half, with nested development/validation selection within each physical realization & Second chronological half & Acquisition jobs \\
Google memory & Samples 05--08 for coordinates; samples 09--12 for $\alpha$ & Samples 13--20 & Acquisition samples \\
Cultivation & First quarter of each released-order basis/depth stream for initial coordinates; second quarter for $\alpha$; coordinates refit on first half & Final half of each stream & Paired 500-shot released-order blocks \\
\bottomrule
\end{tabular}
\end{table}


\section{Logical confidence under model mismatch}
\label{app:statistical-mechanical-decoding}

The main text defines the sector partition functions, exact logical gap, detector-record surprisal, and perturbed Hamiltonian. Here we develop the response terms and the exact-circuit diagnostics beyond those definitions.

\subsection{Linear response of logical confidence}

Expanding Eq.~\eqref{eq:exact-gap-response} about $\epsilon=0$ gives
\begin{equation}
g_\epsilon(d)
=
g_0(d)
+
\epsilon
\left[
\langle V\rangle_{1,d}
-
\langle V\rangle_{0,d}
\right]
+
O(\epsilon^2).
\label{eq:statmech-linear-response}
\end{equation}

Let $\widehat\lambda(d)$ denote the logical sector selected by the frozen decoder and orient the physical confidence toward that decision,
\begin{equation}
h_\epsilon(d)
=
F_{1-\widehat\lambda(d),\epsilon}(d)
-
F_{\widehat\lambda(d),\epsilon}(d).
\end{equation}
Its first-order response is
\begin{equation}
\begin{aligned}
a(d)
&:=
\left.
\frac{\partial h_\epsilon(d)}
{\partial\epsilon}
\right|_{\epsilon=0}\\
&=
\langle V\rangle_{1-\widehat\lambda(d),d}
-
\langle V\rangle_{\widehat\lambda(d),d}.
\end{aligned}
\label{eq:app-oriented-gap-response}
\end{equation}

The response $a(d)$ is the quantity relevant to confidence calibration. A perturbation can make a detector record unusual while shifting both logical sectors equally; such a perturbation changes the absolute record probability but leaves the logical confidence unchanged to first order. Only the sector-asymmetric component of the mismatch contributes to Eq.~\eqref{eq:app-oriented-gap-response}.

For completeness, the second-order expansion is
\begin{equation}
\begin{aligned}
g_\epsilon(d)-g_0(d)
&=
\epsilon
\left[
\langle V\rangle_{1,d}
-
\langle V\rangle_{0,d}
\right]\\
&\quad+
\frac{\epsilon^2}{2}
\left[
\operatorname{Var}_{0,d}(V)
-
\operatorname{Var}_{1,d}(V)
\right]
+
O(\epsilon^3).
\end{aligned}
\end{equation}
Higher orders preserve the same interpretation: only differences between the statistical responses of the logical sectors alter the logical free-energy gap.

\subsection{Exact affine-mixture response}
\label{app:exact-affine-response}

The complete-shot interpolation, oriented confidence $h_\theta(d)$, and first-order response $a(d)$ are defined in Eqs.~\eqref{eq:affine-mixture}--\eqref{eq:exact-mixture-gap-response}. Here we give the identity used to construct the higher-order approximations in Fig.~\ref{fig:gap-correction-residuals}.

Writing
\begin{equation}
\delta_\lambda(d)
=
\frac{q_\lambda^{\mathrm{hot}}(d)}
{q_\lambda(d)}
-1,
\end{equation}
the exact confidence can equivalently be expressed as
\begin{equation}
\begin{aligned}
h_\theta(d)
=
h_0(d)
&+
\log[1+\theta\delta_{\widehat\lambda(d)}(d)]\\
&-
\log[1+\theta\delta_{1-\widehat\lambda(d)}(d)].
\end{aligned}
\end{equation}
The Taylor series used for the residual comparison in Fig.~\ref{fig:gap-correction-residuals}(a) follows directly from this identity.

The resulting conditional-risk response is given in Eq.~\eqref{eq:exact-mixture-risk-response}.

For truncation order $n$, Fig.~\ref{fig:gap-correction-residuals}(a) reports the reference-weighted RMS residual
\begin{equation}
\mathcal E_n(\theta)
=
\left\{
\sum_d q(d)
\left[
h_\theta(d)-h_\theta^{(n)}(d)
\right]^2
\right\}^{1/2},
\label{eq:app-weighted-gap-residual}
\end{equation}
where $q(d)=q_0(d)+q_1(d)$.

For Fig.~\ref{fig:gap-correction-residuals}(b), records are first grouped by either the MWPM gap $\Delta$ or the exact reference gap $|\log[q_0(d)/q_1(d)]|$. Each nonempty gap bin is then divided into 20 conditional $W$ quantiles, and the plotted value is the $q(d)$-weighted mean of $\dot r(d)$ in each joint bin. Because MWPM costs are discrete, a tied score atom that straddles a quantile boundary is fractionally allocated between the adjacent bins; all records in that atom have the same $W$ and response. Repeating the construction with the exact reference gap tests whether the high-$W$ response is merely compensating for the ground-state approximation used by MWPM.

\subsection{Connection to MWPM}

For independent Bernoulli DEM mechanisms, the Hamiltonian and MWPM cost defined in the main text differ by the configuration-independent constant $\kappa_q=-\sum_e\log(1-p_e)$. This constant cancels from the logical gap. The sector partition function retains contributions beyond the minimum-cost configuration:
\begin{equation}
Z_\lambda(d)=e^{-\kappa_q-C_\lambda(d)}
\sum_{E\in\mathcal E_\lambda(d)}
e^{-[C(E)-C_\lambda(d)]}.
\end{equation}
The sum accounts for degenerate minima and higher-cost configurations. Its sector-dependent contribution is the correction neglected by the MWPM ground-state approximation; it is distinct from the physical-model response to $V$ derived above.


\section{Controlled model-mismatch simulations}
\label{app:controlled-drift-methods}

The controlled test uses a Stim rotated surface-code memory with $d=5$, $r=5$, and homogeneous circuit-level Pauli noise at $p_0=0.005$. Each shot draws one quiet or high latent state, shared by all its noisy circuit locations, with high-state probability $\rho=0.1$. The probabilities in Eqs.~\eqref{eq:drifting-circuit-mixture} and \eqref{eq:drifting-circuit-mean} preserve the average location error probability as the burst multiplier $\xi$ changes. We compare a graphlike reference decoder frozen at $p_0$ with a stationary graphlike DEM reconstructed separately at each $\xi$ from the mixture records.

\subsection{Stationary-DEM reconstruction and constrained fit}

Let $d_i\in\{0,1\}$ denote the firing bit of detector $i$. Bulk and boundary mechanism probabilities are inferred from one- and two-detector moments using Eqs.~\eqref{eq:main-bulk-edge-inversion} and \eqref{eq:main-boundary-edge-inversion}. The unconstrained solution is retained when every inferred probability satisfies the graphlike parameterization $0\leq\widehat p_e<1/2$; its first failure defines $\xi_2$.

Beyond $\xi_2$, we use a joint fit constrained to valid mechanism probabilities rather than clipping individual edges. The fit includes all one-detector subsets, the supports associated with two-detector mechanisms, and one alternating half of the connected three- and four-detector subsets. The complementary half of the higher-order subsets is reserved for the diagnostic in Eq.~\eqref{eq:main-higher-order-statistic}.

When all fitted parity moments are positive, the constrained problem uses log-parity coordinates
\begin{equation}
x_e=\log(1-2p_e),
\end{equation}
with $-40\leq x_e\leq0$ and bounded linear least squares. If a fitted moment is nonpositive, the implementation minimizes physical moment residuals directly over $0\leq p_e<1/2$. In both cases the mechanism probabilities are optimized jointly so the result remains a single physical stationary DEM. Numerical arrays and constrained fits use NumPy and SciPy, respectively~\cite{Harris2020NumPy,Virtanen2020SciPy}.

\subsection{Higher-order parity test and stationary null}

To derive the parity prediction in Eq.~\eqref{eq:main-higher-order-prediction}, let $\varepsilon_e\in\{0,1\}$ indicate whether mechanism $e$ fires in a shot, with $\Pr(\varepsilon_e=1)=\widehat p_e$. The detector bits are XORs of the mechanism indicators,
\begin{equation}
  d_i=\bigoplus_{e:\,i\in E_e}\varepsilon_e.
  \label{eq:app-detector-mechanism-xor}
\end{equation}
For a tested detector subset $A$, define the parity variable
\begin{equation}
  P_A=\prod_{i\in A}(1-2d_i).
  \label{eq:app-detector-parity-variable}
\end{equation}
Because XOR adds exponents modulo two, this variable can be written as
\begin{equation}
  P_A=\prod_e(-1)^{\varepsilon_e|A\cap E_e|}.
  \label{eq:app-detector-parity-mechanisms}
\end{equation}
Independence of the mechanism indicators therefore gives
\begin{equation}
  \mathbb E[P_A]
  =\prod_e\left[(1-\widehat p_e)+\widehat p_e(-1)^{|A\cap E_e|}\right].
  \label{eq:app-detector-parity-expectation}
\end{equation}
The factor for mechanism $e$ is $1$ when $|A\cap E_e|$ is even and $1-2\widehat p_e$ when it is odd, giving Eq.~\eqref{eq:main-higher-order-prediction}. The higher-order discrepancy $T_{\mathrm{HO}}$ is defined in Eq.~\eqref{eq:main-higher-order-statistic}.

The null distribution is generated by sampling stationary records from the fitted DEM and repeating the complete fit-and-test procedure. To avoid defining a regime boundary from a single upward fluctuation, we define $\xi_1$ as the first member of the first pair of consecutive scan points whose $T_{\mathrm{HO}}$ values exceed the empirical 95th percentile of the stationary null. Neither this boundary nor $\xi_2$ uses logical outcomes, $W$, or postselection performance.

\subsection{Residual logical information and the high-\texorpdfstring{$W$}{W} tail}

For each frozen decoder, $I(Y;W\mid\Delta)$ is estimated on independent test records, where $Y$ is the logical-failure indicator. Gap bins and the conditional $W$ bins within each gap bin are fixed using calibration data before test labels are accessed. The same estimator is applied to the latent-state oracle $I(Y;Z\mid\Delta)$, which quantifies the logical information that would be available if the quiet/high state were observed directly.

A positive value of $I(Y;W\mid\Delta)$ means that $W$ contains predictive information about logical failure that remains after conditioning on the reported gap. It does not by itself guarantee that one fixed linear score improves every rejection fraction; the postselection test below separately asks whether the chosen likelihood-aware ranking converts that residual information into a lower retained LEP.

We also quantify broadening of the model-incompatibility tail. For each frozen decoder, let $W_{99}$ be the 99th percentile of $W$ under stationary samples from that decoder model. The tail ratio is
\begin{equation}
R_{99}
=
\frac{
\Pr_{\mathrm{mixture}}(W>W_{99})
}{
\Pr_{\mathrm{stationary}}(W>W_{99})
}.
\label{eq:app-tail-ratio}
\end{equation}
This diagnostic asks whether the physical mixture produces more high-cost detector records than the stationary decoder expects; it is intentionally independent of logical outcomes.


\section{Large-distance scaling}
\label{app:burst-scaling}

The exact free-energy calculation in Fig.~\ref{fig:gap-correction-residuals} is necessarily restricted to a small circuit, while the stationary-DEM breakdown in Fig.~\ref{fig:controlled-drift} is designed to expose representational failure rather than distance scaling. We therefore use the circuit-level study in Fig.~\ref{fig:circuit-burst-scaling} to test whether the likelihood-aware advantage persists as both code distance and syndrome volume increase.

\subsection{Simulation and frozen decoder construction}

We use Stim rotated surface-code memory circuits at $d=7$, $11$, and $15$ with the number of syndrome-extraction rounds set to the code distance, $r=d$. The decoder is constructed from the stationary circuit-level model at $p_0=0.005$ and is frozen before the burst arrays are evaluated.

Each physical shot is assigned a whole-shot latent state. With probability $0.9$ every noisy circuit location uses $p_0$, while with probability $0.1$ every location uses the burst probability in Eq.~\eqref{eq:scaling-burst-noise}.
The scaling result in the main text uses $\xi=2$. Unlike the controlled construction of Appendix~\ref{app:controlled-drift-methods}, this mixture is not mean preserving; its purpose is to test whether an elevated-noise subpopulation can be identified increasingly well as the spacetime volume grows.

The $d=7$, $d=11$, and $d=15$ datasets contain 8, 16, and 80 million shots, respectively. The larger $d=15$ sample resolves the substantially lower logical-error probability. The latent state is retained only for the burst-population diagnostic and is not available to either postselection rule.

\subsection{Likelihood-aware score and matched coverage}

Both decoder coordinates are transformed with mid-CDFs computed from independent stationary $p_0=0.005$ reference samples before the score is evaluated. The scaling study fixes $\alpha=0.05$ for all three distances rather than tuning the coefficient independently to each distance or rejection fraction. This deliberately tests transfer of one likelihood-aware correction across increasing spacetime volume. The complete generated burst arrays are then used as evaluation data; no logical labels are used to refit the CDFs, choose $\alpha$, or select an operating point.

For every plotted rejection fraction, gap-only and likelihood-aware selectors are matched to the same retained-shot count. Ties in the discrete decoder scores are resolved independently of the logical outcome. The relative LEP improvement is then computed from the paired retained-error counts.


\section{Hardware decoding and postselection}
\label{app:hardware-methods}

The hardware comparisons require two separations to remain fixed throughout the analysis. Decoder parameters are determined before held-out evaluation, and the likelihood-aware selector is compared with gap-only postselection at the same retained-shot count. This section gives the decoder construction, score fitting, coverage matching, and uncertainty procedures used to enforce those conditions.

\subsection{IBM calibration detector error model}
\label{app:calibration-dem}

The IBM calibration decoder is reconstructed separately for each circuit copy from the calibration snapshot associated with its acquisition. The snapshot supplies qubit relaxation and dephasing times, readout errors and durations, and one- and two-qubit gate errors and durations. These noise parameters are inserted into the circuit before conversion to a Stim detector error model; the primary calibration construction does not add exploratory burst or correlated-error mechanisms.

Idle evolution of duration $t$ is represented by a Pauli-twirled amplitude-damping and dephasing channel. With
\begin{equation}
\eta_{xy}
=
e^{-t/\min(T_2,2T_1)},
\qquad
\eta_z=e^{-t/T_1},
\end{equation}
the inserted channel uses
\begin{equation}
p_X=p_Y=\frac{1-\eta_z}{4},
\qquad
p_Z=\frac{1+\eta_z-2\eta_{xy}}{4}.
\label{eq:calibration-idle-channel}
\end{equation}

Gate and readout errors are inserted using the corresponding reported probabilities and durations, while reset uses the fixed implementation default. Stim then decomposes the resulting circuit-level model into detector mechanisms. The MWPM decoder retains the one- and two-detector graphlike components and their logical labels and converts each mechanism probability to the matching weight
\begin{equation}
w_e=\log\frac{1-p_e}{p_e}.
\end{equation}

\subsection{IBM LEP-optimized decoder}
\label{app:ler-optimized-decoder}

The LEP-optimized decoder keeps the detector and logical connections of the calibration DEM fixed and changes only their rates. It is fitted on the chronological training half; the held-out half is used only after the decoder is frozen.

Fitting has two stages. First, rates are pooled by detector support and adjusted to match the observed distributions of detector patterns in 16 overlapping, 12-detector subsets. The fit minimizes their average smoothed total-variation distance using a deterministic 200,000-shot training subset and at most 80 L-BFGS-B iterations. The resulting DEM supplies the starting rates for the second stage.

The second stage tunes one rate per detector-and-logical signature to minimize full-coverage LEP. For a mechanism with probability $p_e$, its log-rate parameter is $\phi_e=\log[-\tfrac12\log(1-2p_e)]$. Let $L(\boldsymbol\phi)$ be the LEP on a development batch and $L_0$ the LEP of the first-stage decoder on the same batch. Candidates are ranked by
\begin{equation}
\mathcal J(\boldsymbol\phi)
=L(\boldsymbol\phi)+2[L(\boldsymbol\phi)-L_0]_+,
\label{eq:ler-objective}
\end{equation}
where $[x]_+=\max(x,0)$. The penalty disfavors a candidate that performs worse than the starting decoder.

A Gaussian population search runs for 30 iterations, testing 16 candidates per iteration on 20,000-shot development batches and using the best four to update the population. The best candidate from each iteration is also tested on interleaved validation blocks, with at most 40,000 validation shots; the decoder with the lowest validation score is retained. No held-out shot is used in either stage.

\subsection{Google memory decoder priors}
\label{app:google-priors}

The Google analysis uses two supplied detector error models rather than reconstructing a DEM from the held-out data. The first is the RL-optimized prior associated with Google's decoder-prior optimization procedure~\cite{Sivak2024}; the second is the supplied SI1000 prior based on the superconducting-inspired SI1000 circuit-level model~\cite{Gidney2021Honeycomb}. The latter is not a hardware calibration prior and is referred to simply as the SI1000 prior throughout the paper.

Both DEM families are supplied separately for the relevant layout, logical basis, and round count. The analysis freezes the corresponding priors from sample 05 and never refits them using held-out samples 13--20. Earlier Google acquisition samples are used only to construct the likelihood coordinates and select the postselection coefficient.

\subsection{Training-defined likelihood-aware score}
\label{app:training-defined-score}

For a frozen decoder, each detector record produces a complementary gap $\Delta$ and minimum MWPM cost $W$. We use the mid-CDF and likelihood-aware rejection score defined in Eqs.~\eqref{eq:mid-cdf} and \eqref{eq:cdf-score}, with reference or development data assigned to each experiment.

For hardware analyses, candidate values of $\alpha$ and the associated score thresholds are evaluated only on development and validation records. Once an operating point is selected, the empirical CDFs, $\alpha$, and threshold are frozen before held-out evaluation. If $\alpha$ is selected separately at different target coverages, each plotted point should be interpreted as a separately validated operating point rather than as a nested sequence obtained from one universal ranking.

Discrete decoder costs and empirical CDFs produce tied score groups. In the IBM and Google memory analyses, the frozen threshold retains or rejects a complete tied group according to which choice is closest to the target rejection count. The subsequent matched-coverage calculation uses outcome-independent analytic thinning when the two selectors retain different numbers of records. Simulation analyses that require an exact integer count use seeded outcome-independent tie breaking. Logical labels are never used to choose among tied records.

\subsection{IBM uncertainty and estimability}
\label{app:ibm-statistics}

Pooled IBM confidence intervals are obtained from 20,000 paired acquisition-job bootstrap replicates. Jobs are sampled with replacement, all physical copies within a selected job remain together, and the pooled relative improvement is recomputed from matched retained-error counts for both selectors in the same draw. All score transforms, $\alpha$ values, thresholds, decoder parameters, and coverage-matching rules remain frozen. The pointwise bounds are the 2.5th and 97.5th percentiles of the bootstrap distribution; they are not simultaneous bands or intervals for a repeated fitting campaign.

An IBM curve point is displayed only when both matched selectors have at least 50 expected retained logical errors,
\begin{equation}
E_{\mathrm{gap}}\geq50,
\qquad
E_{\mathrm{LA}}\geq50,
\end{equation}
with finite estimates and a positive gap-only denominator. This is an estimability criterion based on retained-error counts, not on the sign of the measured improvement or whether a confidence interval excludes zero.

\subsection{Hardware--DEM minimum-cost diagnostic}
\label{app:dem-sampling-weight}

To compare the hardware $W$ distribution with the frozen decoder model, each IBM physical copy contributes 25,000 held-out QPU shots selected without replacement and 25,000 independent detector records sampled from the corresponding calibration DEM. The same calibration DEM defines the two sector-constrained MWPM problems applied to both sources.

For every detector record we evaluate both logical-sector matchings and retain $W$ as defined in Eq.~\eqref{eq:gap-weight-definitions}. The DEM-sampled record is therefore decoded after sampling; $W$ is not the sum of the particular mechanism weights that happened to generate that simulated record.

For visualization, costs are first accumulated in width-$0.5$ bins and then combined into integer-rounded bins. The 99th-percentile comparison is performed separately within each physical copy before the copy-level ratios are summarized, while the quoted hardware exceedance fraction pools hardware shots relative to their own copy-specific DEM thresholds.


\section{Magic-state cultivation analysis}
\label{app:cultivation-methods}

The cultivation analysis differs from the memory experiments in two ways. Google's native hard acceptance has already removed detected failures during state preparation, and the remaining grafting code is not directly matchable. We therefore construct logical-sector costs from the supplied grafting DEMs and Tesseract decoder, then apply one basis-symmetric likelihood-aware score to the surviving pre-tomography records.

\subsection{Native cultivation acceptance and grafting records}

The released two-cycle and three-cycle datasets are filtered first by Google's native cultivation criterion. The first 16 state-preparation detectors must all be zero. This acceptance rule is never modified by our analysis; likelihood-aware postselection is applied only to records that have already passed it.

For two-cycle cultivation, the postselection features use the complete pre-tomography prefix of 53 detectors, and for three-cycle cultivation they use the corresponding prefix of 80 detectors. Nonzero grafting and error-correction detectors after the first 16 are retained and decoded. Measurements and detectors belonging to the subsequent tomography operation are excluded from the score construction.

\subsection{Projected Tesseract sector costs}

The released $X$- and $Z$-reference detector error models are projected onto the corresponding pre-tomography detector prefix while retaining logical observable $L_0$. To obtain constrained logical-sector costs, one synthetic detector is appended to encode the logical-sector condition. Each observed detector record is then decoded twice with Tesseract~\cite{Tesseract2025}, once with the synthetic detector fixed to sector 0 and once to sector 1.

For each projected reference model $m\in\{X,Z\}$, Tesseract returns the selected error configuration for each constraint and its corresponding cost. We denote these costs by
\begin{equation}
C_{0,m}(d),
\qquad
C_{1,m}(d),
\end{equation}
and define
\begin{equation}
\begin{aligned}
\Delta_m(d)&=|C_{1,m}(d)-C_{0,m}(d)|,\\
W_m(d)&=\min[C_{0,m}(d),C_{1,m}(d)].
\end{aligned}
\label{eq:cultivation-sector-costs}
\end{equation}

Every eligible record is evaluated under both projected reference models regardless of the basis in which its tomography outcome was later measured. These projected sector costs are analysis quantities constructed for the present sensitivity study and should not be identified with Google's released decoder output. The unprojected supplied DEMs were checked independently where possible; the projected models containing logical-only mechanisms are evaluated with Tesseract for the analysis reported here.

\subsection{Basis-symmetric likelihood-aware score}

For each projected reference model $m\in\{X,Z\}$, empirical CDFs are fitted after pooling eligible records from both measurement bases. We define
\begin{equation}
\begin{aligned}
u_m(d) &= 1-\widehat F_{\Delta,m}[\Delta_m(d)], \\
v_m(d) &= \widehat F_{W,m}[W_m(d)].
\end{aligned}
\end{equation}

The two reference models are combined symmetrically through
\begin{equation}
\begin{aligned}
u(d) &= \max[u_X(d),u_Z(d)], \\
v(d) &= \max[v_X(d),v_Z(d)].
\end{aligned}
\end{equation}
and the cultivation rejection score is
\begin{equation}
S_\alpha(d)
=
u(d)+\alpha v(d).
\label{eq:cultivation-score}
\end{equation}
A single $\alpha$, one pooled score, and one pooled threshold are applied to both measurement bases.

The term ``basis-blind'' in the main text refers to this final ranking rule: the score applied to an individual record does not switch between basis-specific postselection policies. Basis labels and validation tomography outcomes are nevertheless required to estimate the validation fidelity used to choose $\alpha$. Held-out tomography outcomes are never used to select $\alpha$, refit the score coordinates, or determine a threshold.

\begin{figure*}[t!]
  \centering
  \includegraphics[width=\textwidth]{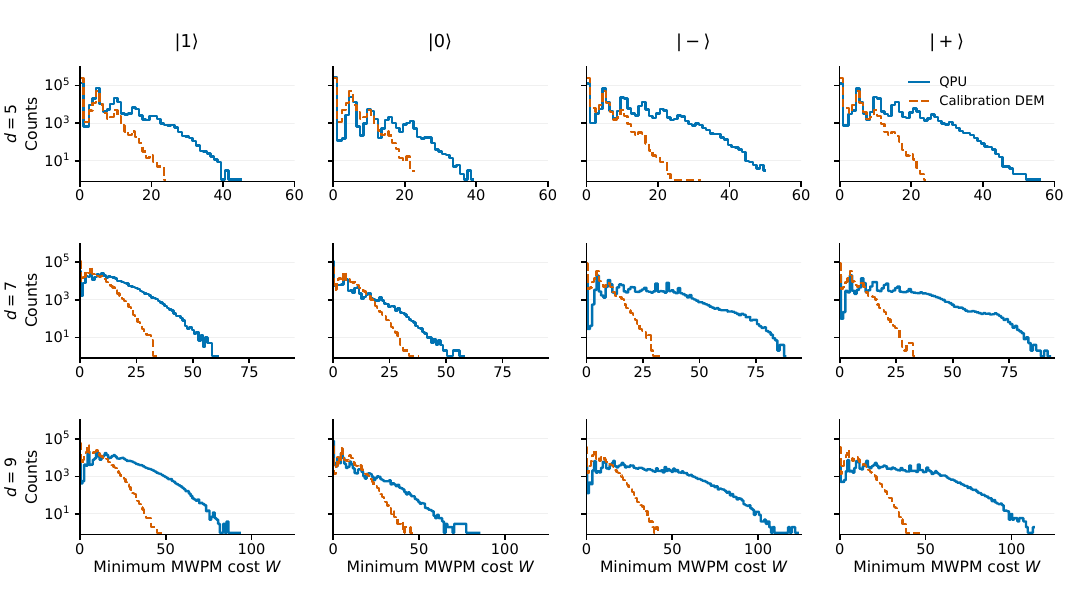}
  \caption{Minimum-cost distributions for every IBM preparation. Rows, from top to bottom, are $d=5$, $7$, and $9$; columns are $|1\rangle$, $|0\rangle$, $|-\rangle$, and $|+\rangle$. Solid blue curves show hardware records and dashed orange curves show calibration-DEM samples. Counts are pooled across realizations and displayed logarithmically. No uncertainty intervals are shown.}
  \label{fig:weight-mismatch-all-states}
\end{figure*}

\subsection{Released-order score selection and held-out evaluation}

After native cultivation acceptance, records are ordered according to the released row order separately within each cultivation depth and measurement basis. The exact eligible counts and frozen boundaries are
\begin{table}[t]
\caption{Released-order cultivation split after Google's native hard acceptance. Intervals are half-open indices within each measurement basis; the release does not provide a single global chronology across bases.}
\label{tab:cultivation-splits}
\centering
\scriptsize
\begin{tabular}{@{}ccrrrr@{}}
\toprule
Cycles & Basis & Eligible & Development & Validation & Held-out \\
\midrule
2 & $X$ & 25,266 & $[0,6316)$ & $[6316,12633)$ & $[12633,25266)$ \\
2 & $Z$ & 28,089 & $[0,7022)$ & $[7022,14044)$ & $[14044,28089)$ \\
3 & $X$ & 33,291 & $[0,8322)$ & $[8322,16645)$ & $[16645,33291)$ \\
3 & $Z$ & 31,186 & $[0,7796)$ & $[7796,15593)$ & $[15593,31186)$ \\
\bottomrule
\end{tabular}
\end{table}

The first quarter is used to construct the CDF coordinates applied during validation, and the second quarter selects $\alpha$. After this choice is frozen, the CDFs are refit using the complete first half and the final half is evaluated. The $X$ and $Z$ streams retain their separate released ordering; they are not interleaved into an artificial global chronology.

\subsection{Target-state fidelity and fixed-rejection comparison}

The tomography quantity used throughout the cultivation analysis is the fidelity with the target magic state,
\begin{equation}
F_T
=
\frac{1}{2}
\left[
1+
\frac{
\langle X\rangle+\langle Z\rangle
}{
\sqrt{2}
}
\right].
\label{eq:cultivation-target-fidelity}
\end{equation}
The reported tomography infidelity is $1-F_T$. This target-state definition is used consistently for validation, the fixed-rejection comparison, the relative infidelity reduction, and the accepted-rate comparison at fixed target fidelity.

At a fixed additional rejection fraction, gap-only and likelihood-aware selectors are compared at matched retained coverage. The relative infidelity reduction is
\begin{equation}
\mathcal I_F
=
1-
\frac{
1-F_{\mathrm{LA}}
}{
1-F_{\mathrm{gap}}
}.
\label{eq:cultivation-infidelity-improvement}
\end{equation}
Because native cultivation acceptance is applied before either selector, the rejection fraction in this comparison refers only to the additional grafting-stage screening introduced after a preparation has already survived Google's hard checks.

\subsection{Accepted rate at fixed target fidelity}

To express the same tradeoff as resource-state yield, the discrete postselection curve is first made monotone by replacing each infidelity value with the cumulative minimum attained up to that rejection level. No extrapolation beyond the validated operating points is allowed.

When a requested target fidelity lies between two adjacent attainable policies, the implementation forms a randomized mixture of those two acceptance policies. The mixture weight is chosen so that the combined accepted ensemble reaches the requested target infidelity exactly, accounting for both the retained counts and retained tomography outcomes rather than visually interpolating the plotted curve. This procedure gives the largest validated retained coverage available at that target.

For target fidelity $F_\star$, let $c_{\mathrm{gap}}(F_\star)$ and $c_{\mathrm{LA}}(F_\star)$ denote the corresponding retained coverages. The accepted-rate increase is
\begin{equation}
\mathcal I_c(F_\star)
=
\frac{
c_{\mathrm{LA}}(F_\star)
}{
c_{\mathrm{gap}}(F_\star)
}
-1.
\label{eq:cultivation-rate-improvement}
\end{equation}
Targets are reported only when at least 95\% of the bootstrap-resampled curves support the requested fidelity without extrapolation.

\subsection{Paired block-bootstrap uncertainty}

Cultivation records exhibit acquisition-order structure, so pointwise uncertainty is estimated with contiguous 500-shot blocks of eligible held-out records rather than independent-shot resampling. The $X$- and $Z$-basis block streams are resampled separately in released order, drawing the original number of blocks within each basis; their retained-shot and tomography-outcome counts are combined to reconstruct $\langle X\rangle$, $\langle Z\rangle$, and fidelity. Gap-only and likelihood-aware policies are evaluated on the same resampled records.

The relative-infidelity-reduction intervals at fixed rejection use 2,000 paired block-bootstrap replicates. The absolute-infidelity bands and matched-fidelity accepted-rate calculation use 4,000 paired replicates; for the latter, each resampled dataset reconstructs the full monotone acceptance--infidelity curve before the target-fidelity comparison is performed. Bounds are the pointwise 2.5th and 97.5th percentiles. The fitted score coordinates, validation-selected $\alpha$, and held-out selection masks remain fixed, so the bootstrap propagates uncertainty through the held-out tomography estimate and, for the matched-fidelity result, through inversion of this curve, but not through model fitting or hyperparameter selection.

\begin{figure}[t]
  \centering
  \includegraphics[width=\columnwidth]{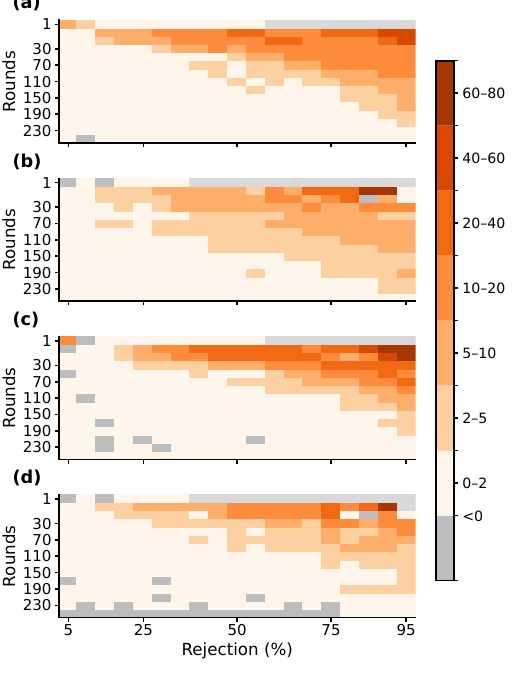}
  \caption{Round-resolved likelihood-aware improvement on Google surface-code data. (a) RL-optimized prior at $d=3$; (b) RL-optimized prior at $d=5$; (c) supplied SI1000 prior at $d=3$; (d) supplied SI1000 prior at $d=5$. Each panel gives the rejection landscape over all available syndrome-extraction durations; colors encode point estimates only, and gray cells denote operating points for which the relative-improvement denominator is undefined.}
  \label{fig:google-improvement-landscape}
\end{figure}

\begin{figure}[t]
  \centering
  \includegraphics[width=0.98\columnwidth]{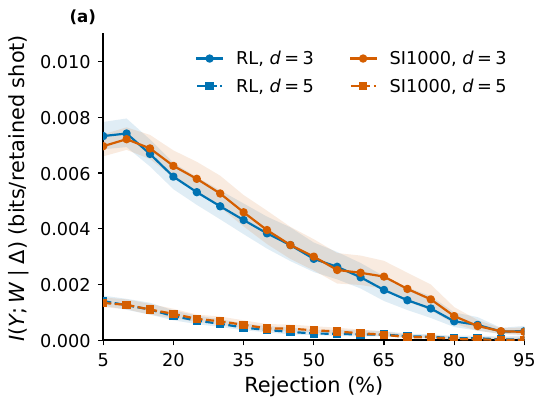}\\[-2pt]
  \includegraphics[width=0.98\columnwidth]{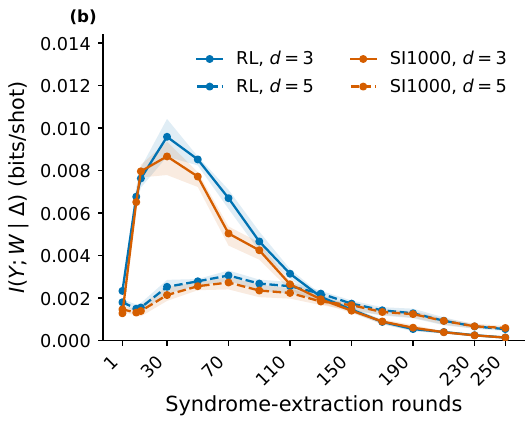}
  \caption{Conditional logical information in the MWPM surprisal proxy on Google data. (a) Bias-corrected $I(Y;W\mid\Delta)$ after gap-only postselection at ten syndrome-extraction rounds, plotted versus rejection fraction for both decoder priors and code distances. (b) Corresponding conditional mutual information across all available round counts before postselection. The eight $\Delta$ bins and four within-$\Delta$ $W$ bins are fit using earlier acquisition samples, while logical labels enter only through the independent held-out evaluation. Shaded regions are pointwise 95\% percentile intervals from 2,000 held-out acquisition-sample bootstrap replicates after a 200-replicate conditional-permutation bias correction.}
  \label{fig:google-cmi-rounds}
\end{figure}


\section{Additional hardware results}
\label{app:supplemental-hardware}

The main text emphasizes pooled held-out performance and a small number of representative operating points. The figures collected here expose state, distance, decoder-prior, and round-count dependence without changing the frozen score-selection procedure. All IBM panels use the final full hardware cohort, including the Kingston acquisition, and all Google panels use the same chronological sample split described in Appendix~\ref{app:common-coverage}.

\subsection{State-resolved minimum-cost tails}
\label{app:weight-mismatch-all-states}

Fig.~\ref{fig:weight-mismatch-all-states} shows that the hardware--DEM discrepancy is not confined to the $|1\rangle$ preparation highlighted in the main text. The shape and extent of the high-$W$ tail vary across the four preparations, with especially broad hardware tails in several $X$-basis panels. Thus, the distribution of detector records as measured by the frozen decoder's minimum cost is preparation dependent. These histograms do not isolate the responsible circuit operations, but they show why the discrepancy should be examined by preparation rather than inferred from one pooled cost distribution.

\subsection{Round-resolved improvement and conditional information}

The full landscape in Fig.~\ref{fig:google-improvement-landscape} separates dependence on code distance, syndrome-extraction duration, rejection fraction, and decoder prior. These panels are descriptive held-out summaries: the score coordinates and $\alpha$ values remain those selected from the earlier development and validation samples.

\begin{figure*}[t!]
  \centering
  \includegraphics[width=\textwidth]{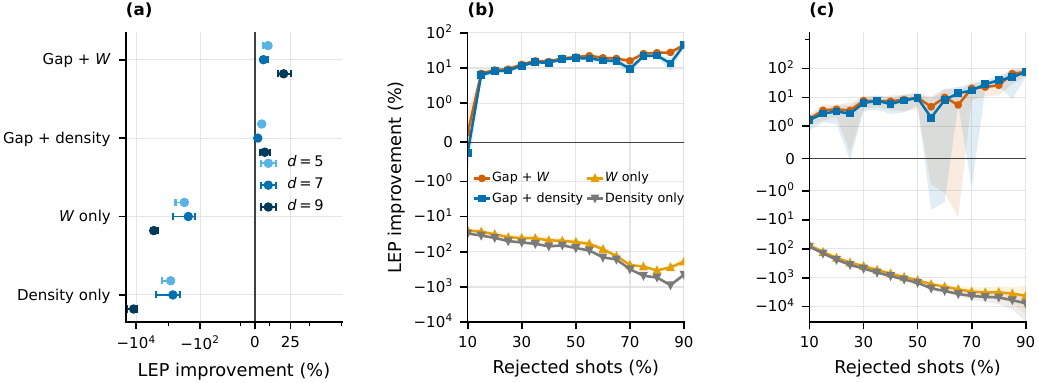}
  \caption{Improvement over gap-only postselection at matched retained coverage. Positive values indicate a lower retained LEP than gap-only selection. (a) IBM results at 10\% nominal rejection using the LEP-optimized decoder, with code distances shown separately. Points and intervals are pooled estimates and pointwise 95\% percentile intervals from paired acquisition-job bootstrap replicates. (b) and (c) Ten-round Google memory results using the RL-optimized prior at $d=3$ and $d=5$, respectively. Curves compare the CDF-normalized gap+$W$ and gap+density scores with $W$-only and density-only selection; bands are pointwise 95\% percentile intervals from paired bootstrap replicates over the eight held-out acquisition samples. The signed symmetric-logarithmic percentage axes are linear between $-1\%$ and $1\%$ and logarithmic outside that interval, retaining visibility of both the positive combined-score effects and the substantially worse single-coordinate controls.}
  \label{fig:method-lift-comparison}
\end{figure*}

At ten rounds, Fig.~\ref{fig:google-improvement-landscape} shows a broad range of rejection fractions with positive point-estimate improvement for both decoder priors, including the distance-$5$ setting emphasized in the main text. The advantage is less uniform at one round and contracts at long durations, where the logical error probability approaches one half and both selectors have less useful ordering to exploit. These features make ten rounds a representative operating point for the main-text coverage curves, while the full duration dependence remains visible here.

Fig.~\ref{fig:google-cmi-rounds} provides a score-independent check of this duration dependence: it tests whether $W$ predicts held-out logical failure after conditioning on the MWPM gap, without assuming that the score in Eq.~\eqref{eq:cdf-score} is optimal.

At ten rounds, Fig.~\ref{fig:google-cmi-rounds}(a) finds conditional information in $W$ for both priors and distances, establishing that the operational gains in Fig.~\ref{fig:google-improvement-landscape} are accompanied by information beyond the reported gap. The information does not peak at ten rounds: Fig.~\ref{fig:google-cmi-rounds}(b) places its maximum near 30 rounds for $d=3$ and 70 rounds for $d=5$, followed by a decline as the memories approach random logical outcomes. The information maximum need not coincide with the best operating point of a particular postselection score. Lower physical error per round could extend the useful range to longer memories if the same model mismatch remains, but these data do not establish that the magnitude of the advantage would increase.


\section{Comparison with related postselection methods}
\label{app:prior-postselection}

Complementary-gap postselection is the baseline for this comparison rather than one of the alternatives introduced here. It already measures logical ambiguity by comparing the best explanations in competing logical sectors. The question tested in this appendix is narrower: whether model-compatibility information improves that baseline, and whether the same held-out ordering can be recovered by other soft-output criteria.

\subsection{Full-cohort comparison protocol}

The final comparison uses the same 163-realization IBM cohort as the primary retained-coverage analysis. Every method is evaluated on the same held-out records for a given frozen decoder, while method-specific hyperparameters are chosen from chronological training and validation data only. No held-out logical label is used to choose a score, threshold, reweighting parameter, or operating point.

We compare gap-only postselection, likelihood-aware postselection using Eq.~\eqref{eq:cdf-score}, $W$ alone, detector density alone, and a CDF-normalized gap-plus-detector-density score. Detector density is the fraction of detector bits equal to one in the record.

Because the methods produce different discrete score distributions, their realized rejection fractions need not coincide exactly. We therefore identify a common retained count within each physical realization and use outcome-independent thinning where necessary. All five selectors are compared only after this coverage matching has been applied.

\FloatBarrier
\bibliography{references}

\end{document}